\documentclass{aa}
\usepackage[utf8]{inputenc}
\usepackage[varg]{txfonts}
\usepackage{graphicx}
\usepackage{wasysym}
\usepackage{natbib,twoopt}
\usepackage[breaklinks=true]{hyperref} 
\usepackage{multirow}
\usepackage{bigints}
\usepackage[table,xcdraw]{xcolor}
\usepackage{booktabs}
\bibpunct{(}{)}{;}{a}{}{,} 
\makeatletter
\newcommandtwoopt{\citeads}[3][][]{\href{http://adsabs.harvard.edu/abs/#3}%
{\def\hyper@linkstart##1##2{}%
\let\hyper@linkend\@empty\citealp[#1][#2]{#3}}}
\newcommandtwoopt{\citepads}[3][][]{\href{http://adsabs.harvard.edu/abs/#3}%
{\def\hyper@linkstart##1##2{}%
\let\hyper@linkend\@empty\citep[#1][#2]{#3}}}
\newcommandtwoopt{\citetads}[3][][]{\href{http://adsabs.harvard.edu/abs/#3}%
{\def\hyper@linkstart##1##2{}%
\let\hyper@linkend\@empty\citet[#1][#2]{#3}}}
\newcommandtwoopt{\citeyearads}[3][][]%
{\href{http://adsabs.harvard.edu/abs/#3}
{\def\hyper@linkstart##1##2{}%
\let\hyper@linkend\@empty\citeyear[#1][#2]{#3}}}
\newcommand{\wat}{\mathrm{H}_2\mathrm{O}} 
\newcommand{\car}{\mathrm{CO}_2} 
\newcommand{\nit}{\mathrm{N}_2} 

\makeatother

\begin{document}

\title{When the core ate the atmosphere}

\author{Vera~Sparrman\inst{\ref{Copenhagen}}\and Anders~Johansen\inst{\ref{Copenhagen}, \ref{Lund}}}

\institute{Center for Star and Planet Formation, Globe Institute, University of Copenhagen, Øster Voldgade 5-7, DK-1350 Copenhagen, Denmark \label{Copenhagen} \and Lund Observatory, Department of Astronomy and Theoretical Physics, Lund University, Lund, Sweden \label{Lund}}

\abstract{The initial volatile element inventories in a planet's atmosphere, mantle, and core are established by delivery and transport during planetary growth. Conversely, the volatile abundances of Earth's planetary reservoirs record its accretion pathway. We model here volatile transport during planet formation for two different delivery scenarios, via pebble and planetesimal accretion, to predict the volatile inventories of proto-Earth at the end of its main accretion phase. 
We find that the distributions of volatiles between the atmosphere, mantle, and core exhibit strong sensitivity to volatile delivery timing, metal-silicate partition coefficients, atmosphere-magma solubilities, and planet mass. In particular, the early volatile delivery expected in pebble accretion scenarios substantially enhances core sequestration of hydrogen and carbon relative to continuous delivery during incremental planetesimal accretion.
With our fiducial partition coefficients, the pebble accretion model produces a 760 bar carbon-dominated atmosphere with 39 bar of hydrogen and water and a few bars of nitrogen, consistent with volatile concentration estimates for bulk silicate Earth including the atmosphere. 
The corresponding planetesimal accretion model produces a more massive ($\sim$$1350$ bar) atmosphere with similar composition. 
The core stores $\sim$$70$\% of both the H and the C budget in the early-delivery scenario, but $\lesssim$$50$\% in the corresponding continuous delivery scenario. 
Overall our results demonstrate that substantial amounts of volatiles are  sequestered into rocky planet cores during planet formation and that the volatile inventory of bulk silicate Earth was likely established already during the magma ocean phase of planet formation.
}

\keywords{Earth / planets and satellites: composition /  planets and satellites: formation / planets and satellites: interiors / planets and satellites: terrestrial planets}

\maketitle

\section{Introduction} \label{Introduction}

To date, more than 6000 exoplanets have been detected. However, size and orbit biases in the dominant detection methods favor the detection of planets larger and hotter than the terrestrial planets of the solar system, such as super-Earths, sub-Neptunes and hot Jupiters. The James Webb Space Telescope \citep{JWST2006} has recently advanced detection methods to the possibility of characterizing atmospheres of rocky exoplanets down to the size of Earth \citep[e.g.][]{Lichtenberg2025}. 
Thick $\car$-dominated atmospheres have generally been excluded for the rocky planets targeted so far based on their lack of atmospheric heat redistribution, which may hint at unexpectedly rapid atmospheric loss or that carbon is trapped in the planetary interiors \citep[e.g.][]{Kreidberg2025}. 

In contrast to exoplanets, the atmospheres of the rocky planets in the Solar System are well characterized. Earth's atmosphere is unique among the rocky planets in the Solar System by being poor in CO$_2$ and rich in O$_2$. Although the atmospheres of Mars and Venus differ in pressure by several orders of magnitude, they have remarkably similar compositions. Carbon dioxide dominates their atmospheres (95\%, 96.5\% for Mars and Venus respectively), accompanied by a few percent (2.8\%, 3.5\%) of molecular nitrogen \citep[e.g.][]{Lecuyer2000, Basilevsky2003, Franz2017, Way2020}. The difference between the current atmospheres of Venus, Earth and Mars may stem from a myriad of processes including volatile cycling between mantle and atmosphere \citep{Plank2019, Graham2020, Honing2021}, a difference in tectonic modes \citep[e.g.][]{Honing2019, Cawood2022, Weller2020, Marshall2024}, atmospheric escape \citep[e.g.][]{Kasting1983, Way2020}, and interactions with the biosphere in Earth's case \citep[e.g.][]{Archean2020, Knoll2017, Wang2025}. The terrestrial planets may already have differed in their initial chemical inventories endowed to them at the formation stage. Disentangling these scenarios could inform us whether Earth's atmosphere was always unique, like today, or whether rocky planet formation naturally produced similar atmospheres for both Venus, Earth and Mars. 

The atmosphere is a key component in the study of life. First, for a given incident stellar flux, some atmospheres may produce a temperate surface that can host life, while other atmospheres may not \citep[e.g.][]{Kopp2013, Kasting1993, Methane2018}. Second, certain atmospheric species, such as O$_2$, PH$_3$ \citep[][]{Greaves2021phosphine}, or more recently dimethylsulfide \citep[][]{Madhusudhan2023}, have been proposed as biomarkers that indicate the presence of life, although such inferences tend to be inconclusive in part due to the difficulty in ruling out alternative synthesis processes \citep[e.g.][]{Greaves2021addendum, Bains2021, Clements2024}, alternative composition fits to spectra \citep[e.g.][]{Wogan2024, Shorttle2024}, or overestimation of the detection certainty \citep[e.g.][]{Kipping2025overestimate}. Third, biomolecules within cells consist mainly of the elements H, C, N which are atmophile in their oxidized forms ($\wat$, $\car$,  N$_2$). 
The atmospheric composition of a newly-formed planet therefore informs us both about the availability of liquid surface water and about the H, C and N budgets necessary for the synthesis of biomolecules. 

Life emerged early in the geological history of Earth, during a time for which the atmosphere is poorly constrained.
Fossil records of life on Earth extend to near the beginning of the Archean $\sim$$3.7$ Gyr ago \citep[e.g.][]{Stromatolites2007, Allwood2016, Nutman2016}. Few rocks remain from the preceding Hadean eon, which lasted from the formation of Earth $\sim$$4.567$ Gyr ago \citep{Connelly2012}, to approximately 4.0 Gyr ago \citep{Knoll2017}, limiting the possibility of finding fossil evidence of an earlier origin of life or inferring the surface conditions of early Earth. Recent phylogenetic analysis dates the last universal common ancestor to $\sim$$4.2$ Gyr ago \citep{Moody2024}, suggesting an even earlier origin of life. Once Earth's physical and chemical conditions were conducive to life, both fossils and phylogenetic analysis thus suggest that life's emergence was quick \citep{Kipping2020, Kipping2025abio}.
If life is scarce in the universe, which could be a solution to Fermi's paradox \citep[e.g.][]{Hart1975, Gray2015, Sandberg2018}, the inference of fast abiogenesis therefore suggests that early Earth analogues could be rare in our galaxy. This connection emphasizes the need to understand the prevalence of early Earth conditions and, by extension, the planetary formation processes that produced those conditions. 

Volatile budgets relevant for life and atmospheric composition are established during planet growth. Planets grow from planetary embryos that accrete material from the surrounding protoplanetary disk during the formation of the stellar system \citep[e.g.][]{Safronov1972}. 
In the protoplanetary disk, dust grains coagulate and then grow through collisions into mm-cm sized pebbles, but increasing collision speeds make larger solids prone to fragmentation \citep[e.g.][]{Guttler2010, Seizinger2013}
Once solids exceed sizes of a few hundred meters, self-gravity inhibits collisional dispersion \citep[e.g.][]{Benz1999}; such gravitationally bound bodies are referred to as planetesimals. Overcoming the growth barrier between pebbles and planetesimals likely requires local clumping of pebbles, for example by gravitational instability in the disk midplane \citep[e.g.][]{Youdin2002} or from the streaming instability \citep[e.g.][]{Youdin2005, Johansen2007, Johansen2012}.

The main formation pathways for planets are planetesimal accretion, pebble accretion and giant impacts. Classical planetesimal accretion models grow planetary embryos by sweep-up of smaller planetesimals \citep{Safronov1972, Morbidelli2022, Batygin2023}, but such models struggle to grow the planetary core of the gas giants of the solar system before the protoplanetary disk disperses, unless the planetesimals are very small ($\sim$$100$ meters) \citep[e.g.][]{Fortier2013, Johansen2019}. Pebble accretion has developed over the last two decades as an alternate pathway for planet formation with a higher accretion rate in the outer disk, capable of forming gas giants cores within the lifetime of the protoplanetary disk \citep{Ormel2010, Lambrechts2012, Ida2016, Visser2016, Johansen2017, Bitsch2019}. Both planetesimal accretion and pebble accretion could have contributed to the growth of the terrestrial planets of the Solar System in the disk lifetime \citep{Johansen2021, Olson2023, Wang2026}. 

During accretion, protoplanets host global magma oceans that enable the transport of volatiles between planetary reservoirs. Planetary growth in the protoplanetary disk phase supplies an enormous accretion luminosity that raises temperatures above the silicate melting point throughout the crust and mantle \citep[e.g.][]{Rubie2015, Steinmeyer2023, Anatomy2}.
The transport of volatiles from the magma ocean to the core is therefore a natural consequence of planetary differentiation and core formation during the magma ocean stage \citep[e.g.][]{Nimmo2015, Fegley2020, Olson2022, Anatomy3}. At the high temperatures prevailing above the magma ocean, surface volatiles are entirely carried by gaseous species. 

The efficiency of volatile transport between the atmosphere, mantle and core of growing protoplanets depends on the solubilities and metal-silicate partition coefficients of volatiles. 
Experimental studies of volatile partition coefficients are beginning to reach pressures and temperatures sufficient to probe volatile partitioning under magma ocean conditions for Earth-mass planets \citep[e.g.][]{Grewal2019, Fischer2020, Grewal2021, Speelmanns2019, Blanchard2022, Huang2024, Luo2024}. These experimental results inform models of volatile transport between planetary reservoirs, which are crucial for our understanding of the volatile inventories of young rocky planets \citep[e.g.][]{Ortenzi2020, Sossi2020, Bower2022, Gaillard2022outgassing, Sossi2023, Anatomy3, Gu2024, Rogers2025, Modirrousta2025}. 

In this study, we analyze the transport of the volatiles H, C, and N during the growth of rocky planets to constrain their volatile inventories at the end of accretion. Notably, understanding the composition of Earth's early atmosphere is crucial for constraining the climate and the chemical conditions during the origin of life. We combine the effects of volatile sedimentation during core-formation with magma ocean interactions with the outgassed atmosphere. To properly account for the pressure and temperature dependence of metal-silicate partition coefficients, we radially resolve the magma ocean as the planet grows. We focus in particular on volatile delivery timing, which differs between pebble accretion and planetesimal accretion scenarios, and quantify how this difference influences the resulting volatile inventories. 

Our paper is structured as follows. In Sect. \ref{sec:methods}, we describe our model for volatile transport during the growth of rocky planets. In Sect. \ref{sec:results} we present the results of several simulations of proto-Earth analogues formed by pebble accretion or planetesimal accretion. In Sect. \ref{sec:discussion} we contrast volatile inventories resulting from different accretion scenarios and model assumptions, and assess their sensitivities to such choices. We also compare the volatile inventories to estimates of the volatile content of Earth's planetary reservoirs. In addition, we discuss volatile exchange after the primary accretion phase and how our results extrapolate to Venus and more massive terrestrial planets. Finally, in Sect. \ref{sec:conclusion}, we summarize our findings, highlighting the importance of volatile delivery timing and its implications for constraining Earth's formation pathway. 

\section{Methods} \label{sec:methods}
\begin{figure}[]
\resizebox{\hsize}{!}{\includegraphics{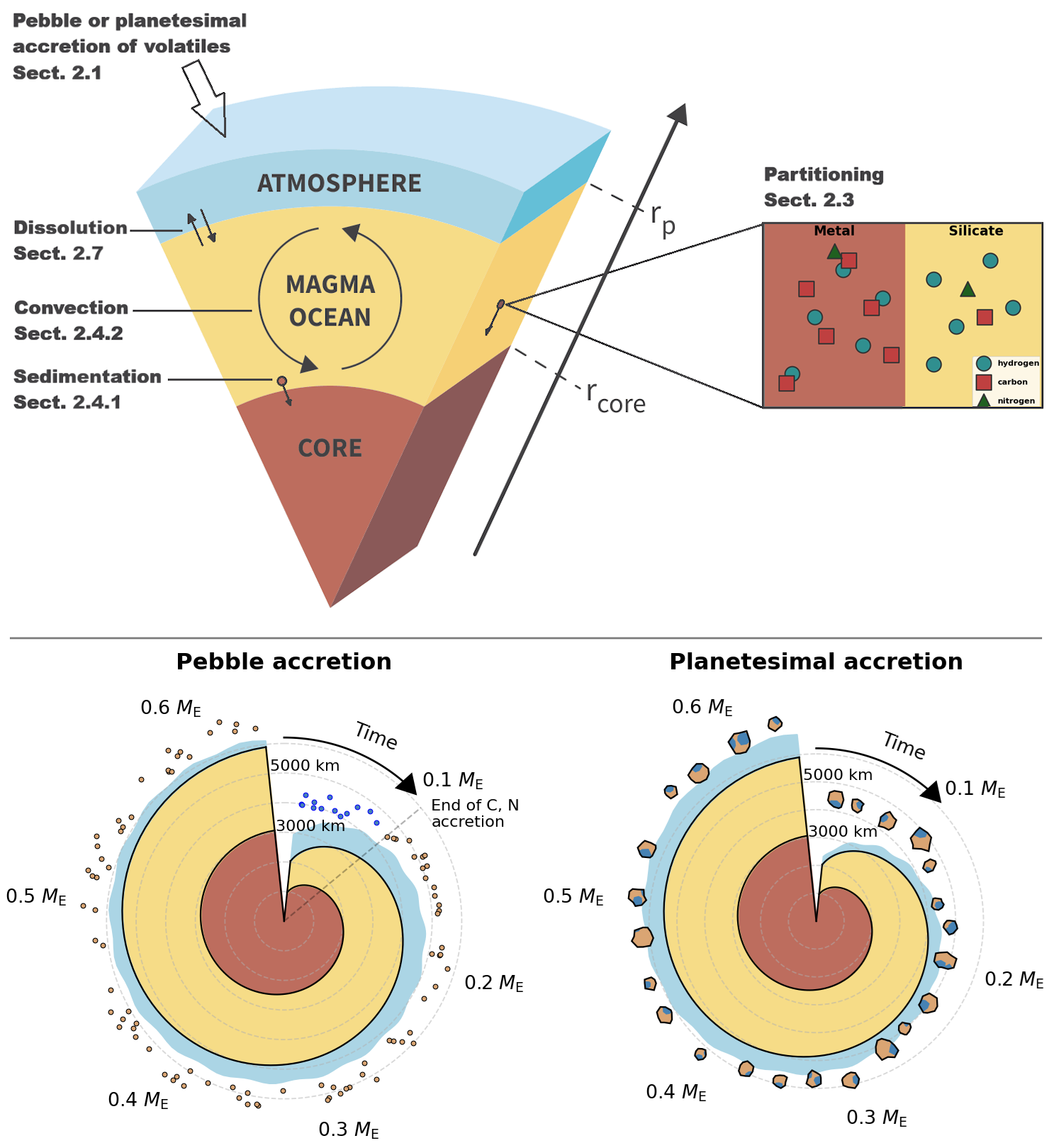}}
\caption{Volatile transport model during the magma ocean stage of protoplanetary growth. \textit{Top:} Overview of volatile transport processes between and within the planetary reservoirs: the atmosphere, the magma ocean, and the core. \textit{Bottom:} Growth of proto-Earth by pebble and planetesimal accretion. In the pebble accretion scenario, volatile delivery of H occurs before $M_\text{p} = 0.02 \, M_\text{E}$ and C and N occurs before $M_\text{p} = 0.1 \, M_\text{E}$ (blue). Pebbles accreted after are volatile-poor (orange). In the planetesimal accretion scenario, the planetesimals (blue and orange) contain a constant volatile fraction during the growth of proto-Earth.}
\label{snailfig}
\end{figure}

In this section, we describe our model for the growth of rocky (proto-)planets within a protoplanetary disk. In Sect. \ref{sec:planetgrowth} we describe the planet growth rate and the accretion of volatiles. Section \ref{sec:interior} presents our planetary interior model. Section \ref{sec:partitioncoef} details how volatiles partition between metal and silicates in the magma ocean, while Sect. \ref{sec:magmaoceantransport} describes the resulting volatile transport. Finally, Sects. \ref{sec:oxygenfugacity}, \ref{sec:speciation} and \ref{sec:dissolution} describe the gas-gas and gas-melt interactions of volatiles. Figure \ref{snailfig} depicts a schematic overview of our model.

\subsection{Planet growth} \label{sec:planetgrowth}
We calculate the growth of proto-Earth starting at a planet mass of $M_\text{p} = 0.02 \, M_\text{E}$ after the mantle is fully molten and the first metal-rich core has separated from the silicates \citep[e.g.][]{Olson2022, Anatomy2}. In this phase, we assume that planet growth proceeds via either pebble accretion or planetesimal accretion until reaching $M_\text{p} = 0.64\,M_\text{E}$ (\citealt{Wang2026}). We choose a stopping mass $<$$1\,M_\text{E}$ to represent the mass of proto-Earth before the Moon-forming giant impact and any additional planetesimal impacts that occur after the end of the disk phase \citep[e.g.][]{Olson2023, Wang2026}. Even classical \textit{N}-body simulations of terrestrial planet formation predict significant planetary growth during the first 10 Myrs of accretion \citep[e.g.][]{Kominami2002, Woo2022}, followed by a late accretion phase consisting of giant impacts \citep[e.g.][]{Quintana2016, Woo2023}. 

We prescribe a simplified planet growth rate with a constant accretion timescale $T_\text{acc}$ as
\begin{equation}
    \dot{M}_\text{p} = \frac{M_\text{p}}{T_\text{acc}}.
\end{equation}
The accreted mass is divided into $\sim$$70\%$ silicates and $\sim$$30\%$ iron-nickel, to mimic the core-mantle mass ratio of Earth \citep{Loper2003}. We also consider the accretion of volatiles, which in total represent $\lesssim$1\% of the mass. We focus on the volatiles H, C, and N because they are fundamental volatiles for life on Earth and Earth's atmosphere. The O content of the volatiles is implicit once they are accreted, mediated via the IW buffer in the magma ocean \citep{Hirschmann2021} (see Sect. \ref{sec:oxygenfugacity}).

We model two end member cases for volatile delivery: pebble accretion and planetesimal accretion. As rocky planets grow, the temperature of their surrounding envelope increases, which strips incoming pebbles of their volatiles by sublimation \citep[e.g.][]{Alibert2017, Brouwers2018, Steinmeyer2023, Anatomy2, Wang2023}. In our pebble accretion case, we therefore assume that envelope processing stops the accretion of each volatile at specific mass thresholds, to reflect the heating of the pebbles during their descent to the magma ocean. In this case, hydrogen accretion (carried by $\wat$) stops at the initial mass ($M_p \approx 0.02\, M_\text{E}$) while carbon and nitrogen accretion (carred by organics and graphite) stops around Mars mass ($M_p \approx 0.1\, M_\text{E}$) protoplanets \citep{Anatomy3}. If Earth instead accreted primarily larger bodies, volatiles in their interior would be protected from sublimation and continue to be delivered irrespective of the envelope temperature. In the planetesimal accretion case, we therefore model volatile delivery as a continuous process that proceeds over the entire accretion history of the planet. Some recent studies report sufficient hydrogen abundances in the enstatite chondrites, a class of meteorites that likely represent conditions interior to the water ice line in the solar protoplanetary disk, to account for the bulk hydrogen content of Earth \citep{Piani2020, Thomassin2023}. However, \citet{Gray2026} estimate that enstatites could account for at most one quarter of Earth's water, based on analysis of water in minerals that are less sensitive to terrestrial contamination.

In both the pebble and the planetesimal cases, we fix the total volatile budget to match that inferred for Earth. The total volatile budget is the sum of the volatiles in bulk silicate Earth (BSE) and in the core. Table \ref{tab:volatilebudget} lists the BSE and core estimates we use to calculate the volatile budget. We adopt estimates of the BSE concentrations from \citet{Hirschmann2018} for $\wat$, C, and N. We calculate the core content of $\wat$ based on an inferred core hydrogen abundance equivalent to $0.63$--$3.24$ wt\% $\wat$ or $9$--$45$ Earth oceans \citep{Huang2026}. The upper limit is approximately $\sim$$50\%$ higher than previous seismic and geophysical estimates of hydrogen in Earth's core \citep[e.g.][]{Hirose2021}, and exceeds the $\sim$$7$ Earth oceans predicted to accrete by a pebble accretion model of proto-Earth \citep{Anatomy3}. We explore both the lower and upper limits in our simulations, but slightly favor the lower limit case given closer alignment to these other estimates. For C, we adopt an estimate of $0.3$ wt\% in the core from a compositional match to seismic measurements of the shear wave speed in the inner core \citep{Chen2014}. 

\begin{table}[]
\caption{Estimates of the volatile inventories of Earth.}
\label{tab:volatilebudget} 
\centering
\begin{tabular}{llll}
\rowcolor[HTML]{CBCEFB} \bottomrule
Reservoir & $\wat$ & C & N
\\ \hline 
BSE\tablefootmark{a} [ppmw] & 710 & 140 & 2.8 \\ 
Core [ppmw] & $(6300, 32400)$\tablefootmark{b} & 3000 \tablefootmark{c} & 0 \\ 
Total [$10^{21} \,$kg] & $(15.4, 64.7)$ & 6.3 & 0.011 \\
\toprule
\end{tabular}
\tablefoot{Bulk silicate Earth and core concentrations of volatiles comprise the total volatile budget of our proto-Earth model. We test lower and upper bounds of the hydrogen content of the core (numbers in parentheses), given significant uncertainties. \tablefoottext{a}{Bulk silicate Earth concentrations retrieved from \citet{Hirschmann2018}.} \tablefoottext{b}{Core H content from \citet{Huang2026} renormalized to $\wat$ mass concentration.} \tablefoottext{c}{Core C content retrieved from \citet{Chen2014}.} 
}
\end{table}

\subsection{Interior structure} \label{sec:interior}

We model our planets as spherically symmetric bodies and describe their magma ocean with the structure variables pressure $p$, temperature $T$, and density $\rho$, as functions of radial distance $r$, from the center of the planet. We model $p$ and $T$ as solutions to a system of differential equations that describe hydrostatic equilibrium and convective heat transfer \citep[e.g.][]{Abe1997, Valencia2007},
\begin{align}
    \frac{dp}{dr} &= -\rho g, \label{eq:dpdr} \\
    \frac{dT}{dr} &= -\frac{\alpha_\text{melt} \, T \, g}{C_p}, \label{eq:dTdr} 
\end{align}
where $g$ denotes the local acceleration due to gravity, calculated as
\begin{equation}
    g(r) = \frac{G M_\text{enc}(r)}{r^2}\,.
\end{equation}
Here $M_\text{enc}(r)$ is the mass enclosed inside $r$, which includes the core mass. We compute the core radius by assuming Earth's average core density of 10480 kg/m$^3$ \citep{Loper2003}.
We approximate $C_p$ of the melt using the specific heat capacity of MgSiO$_3$ from the condensed phase thermochemistry chapter in the NIST chemistry WebBook \citep{Chase1998, NISTWebBook}. 
Equation \ref{eq:dTdr} assumes an adiabatic temperature gradient since magma oceans are strongly convective \citep[e.g.][]{Rubie2003, Dorn2018, Ortenzi2020, Anatomy1, Thomas2024}.
We integrate Eqs.~\ref{eq:dpdr}, \ref{eq:dTdr}, to compute $p$ and $T$,
\begin{align}
    & p(r) = p_\text{surf} + \int_r^{r_\text{surf}} g(r')\rho(r')  \, dr', \\
    & T(r) = T_\text{surf} \, \exp{ \left( - \int_r^{r_\text{surf}} \frac{\alpha_\text{melt}(r') \, g(r')}{C_p(r')} \, dr' \right)} \label{eq:Tintegral}.
\end{align}
To compute the melt density $\rho$, we construct a 2D piecewise-linear interpolation with respect to $p, T$ from the equation of state results for pyrolite in \citet{Caracas2024}. We also linearly interpolate the volumetric thermal expansion coefficient $\alpha_\text{melt}$ from \citet{Caracas2024} using its definition,
\begin{align}
    &\alpha_\text{melt} = - \frac{1}{\rho} \left( \frac{\partial \rho}{\partial T} \right)_p.
\end{align} 

Each structure variable depends on the other structure variables. Therefore, to find a self-consistent structure of the interior for each growth time step, we discretize the magma ocean into equal width layers and compute $p$, $T$ and $\rho$ in that order iteratively until convergence. The width of each layer, and therefore the total radius of the planet $r_\text{surf}$, changes dynamically with $\rho$ to conserve the total mass for all iterations in a single time step. 

We compute $p_\text{surf}$ using the mass of the atmosphere at a given time step (Eq.~\ref{eq:thinatmosphere}). We fix $T_\text{surf} = 3000\,\text{K}$ \citep[e.g.][]{Anatomy2, Steinmeyer2024}. Surface temperatures may affect volatile dissolution (Sect. \ref{sec:dissolution}) and, because the surface temperature sets the temperature of the magma ocean adiabat (Eq. \ref{eq:Tintegral}), the volatile partition coefficients (Sect. \ref{sec:partitioncoef}). 

\begin{figure*}[!t]
\centering
\includegraphics[width=17.5cm]{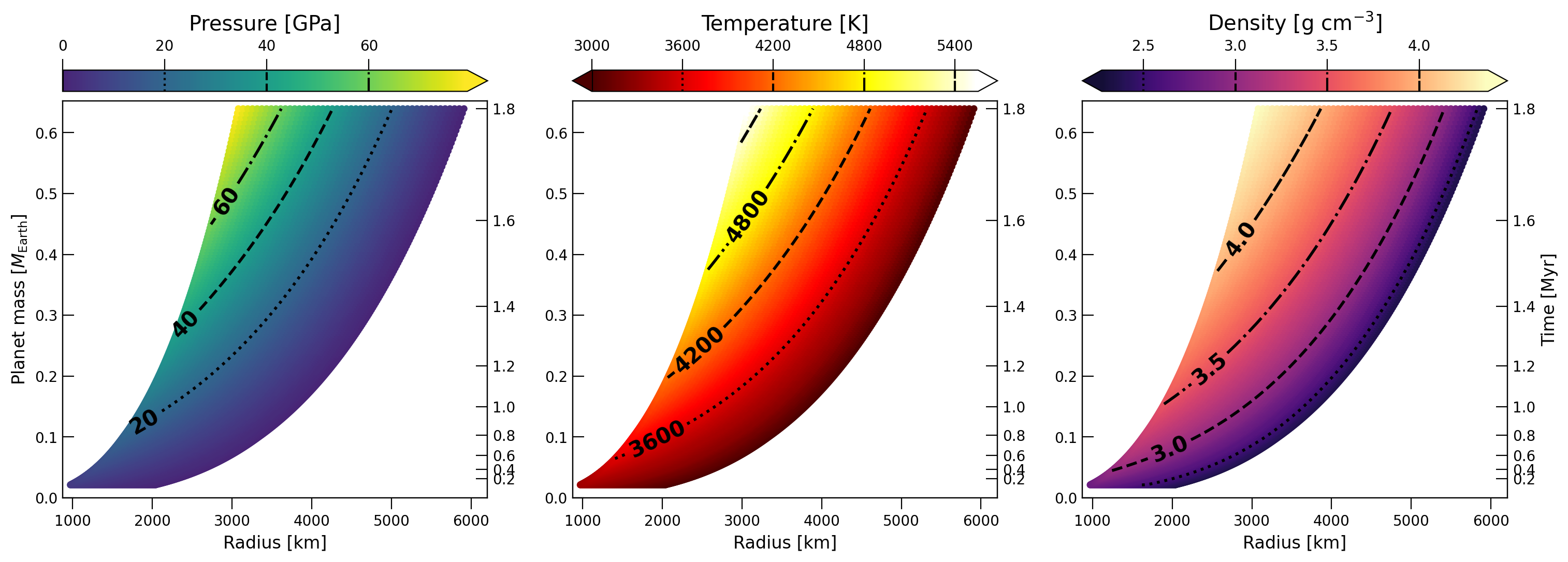}
\caption{Internal structure variables  $(p, T, \rho)$ at every depth of the magma ocean as the planet grows from $0.02 \, M_\text{E}$ to $0.64 \, M_\text{E}$. Horizontal cross sections describe the $(p, T, \rho)$-profiles at a point in time/at a specific planet mass. The left edge of the funnel shape represents the core-mantle boundary and the right edge represents the top of the magma ocean. At any point in time, $(p, T, \rho)$ all increase nearly linearly with depth in the magma ocean.}
\label{widestructure}
\end{figure*}

Fig. \ref{widestructure} shows the pressure, temperature, and density of the rocky planet as it grows from $0.02 \, M_\text{E}$ to $0.64 \, M_\text{E}$. Pressure varies more throughout the magma ocean than temperature or density. Whether partition coefficients depend primarily on $p$ or $T$ therefore matters especially for partitioning near the surface of the magma ocean. Additionally, while the surface conditions remain constant, the $p$, $T$ and $\rho$ at the core-mantle boundary all increase with the mass of the planet. Therefore, the partitioning affinity of volatiles in the deep magma ocean will change as the planet grows.

\subsection{Partition coefficients} \label{sec:partitioncoef}
\begin{figure}[]
\resizebox{\hsize}{!}{\includegraphics{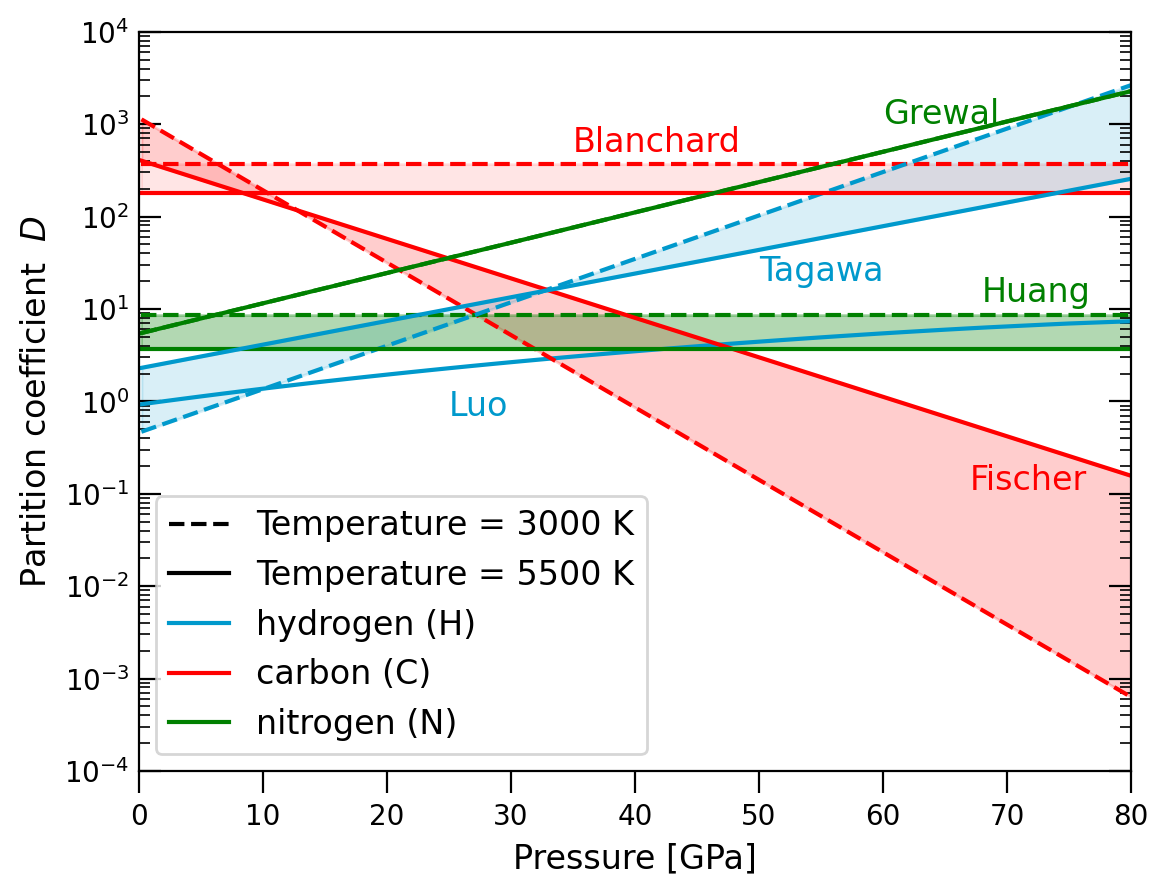}}
\caption{Metal-silicate partition coefficient fits for H, C, N as a function of $p$ and $T$ at an oxiditation level $\text{IW}$$-2$ (see Sect. \ref{sec:oxygenfugacity}). Fits are retrieved from molecular dynamics simulations, H: \citep{Luo2024}, or diamond anvil cell experiments, H: \citep[Eq. 10]{Tagawa2021}, C: \citep{Fischer2020, Blanchard2022}; N: \citep{Grewal2019, Huang2024}. The dashed and the solid line shows the partition coefficient of each fit evaluated at 3000 and 5500 K respectively.}
\label{partitioncoef}
\end{figure}

Volatiles in the magma ocean partition between the silicate magma and the sinking metal droplets according to the local partition coefficient, $D_i$, of the volatile of index $i$, 
\begin{equation}
    D_i = \frac{C_{i}^\text{(met)}}{C_{i}^\text{(sil)}} \label{eq:partitioncoef}
\end{equation}
where $C_{i}^\text{(met)}$ and $C_{i}^\text{(sil)}$ are the equilibrium mass concentrations of volatile $i$ in the metal droplets and the silicate magma respectively \citep[e.g.][]{Luo2024, Fischer2020, Grewal2019, Blanchard2022, Tsutsumi2025, Huang2024, Anatomy3}. To study the high pressures and temperatures of the deep magma ocean, we rely on results of diamond anvil cell experiments and molecular dynamics simulations. Figure \ref{partitioncoef} shows the local partition coefficients of $\wat$ \citep{Tagawa2021, Luo2024}, C \citep{Fischer2020, Blanchard2022}, and N \citep{Grewal2019} parametrized as functions of $p$ and $T$. We describe in more details the choice of partition coefficient sources in Sects. \ref{sec:hydrogen}, \ref{sec:carbon}, and \ref{sec:nitrogen}.

Strikingly, Fig. \ref{partitioncoef} shows that all three volatile elements considered here have partition coefficient fits with several orders of magnitude of tension between different experimental and theoretical approaches under deep magma ocean conditions. The partition coefficient fits can depend on both pressure and temperature, with both pressure and temperature increasing monotonically with depth in the magma ocean (Fig. \ref{widestructure}). 
The pressure varies more throughout the magma ocean than the temperature. Studies that find pressure-dependent partition coefficient fits \citep[e.g.][]{Fischer2020, Grewal2019} therefore result in partition coefficients that vary more throughout the magma ocean than studies that find pressure-independent partition coefficient fits \citep[e.g.][]{Blanchard2022, Huang2024}. Although the pressure-independent partition coefficient fits appear nearly constant compared to the pressure-dependent partition coefficients, the partition coefficients of \citet{Blanchard2022} and \citet{Huang2024} still vary by a factor of 2-3 throughout the magma ocean.

\subsubsection{Hydrogen} \label{sec:hydrogen}
Molecular dynamics simulations have been used to study the hydrogen partition coefficient for the full range of thermodynamic conditions of the magma ocean \citep[][]{Yuan2020, Li2020, Luo2024}. In the $<$$20$ GPa pressure regime previously probed experimentally \citep[e.g.][]{Malavergne2019, Clesi2018}, the simulations generally overlap experimental results \citep{Li2020, Luo2024}. Experiments and molecular dynamics simulations suggest that hydrogen becomes increasingly siderophile at higher pressures. \citet{Yuan2020} and \citet{Li2020} both found a continuation of the experimental trend and concluded that the partition coefficient increases by an order of magnitude between 20 and 40 GPa. \citet[][]{Yuan2020} first predicted a further continuation of the trend even beyond 100 GPa. However, both \citet{Li2020} and \citet{Luo2024} observe that the increase in the partition coefficient tapers off before $\sim$$100$ GPa. Water has also been experimentally measured in a diamond anvil cell by \citet{Tagawa2021} who found highly siderophile partitioning from 30 to 60 GPa.

We choose to compare in this study the hydrogen (water) partition coefficient fits of \citet{Luo2024} and \citet{Tagawa2021}. We choose the partition coefficient of \citet{Luo2024} to represent our fiducial model and explore \citet{Tagawa2021} to probe a more siderophile alternative where more water enters the core.

The partition coefficient of \citet{Luo2024} depends on pressure and the water fraction in the metal itself, but not on the temperature (Fig. \ref{partitioncoef}). \citet{Li2020} have also noted the dependence on the water concentration in the metal. Since the partition coefficient both determines (Eq. \ref{eq:partitioncoef}) and is a function of the water fraction in the metal \citep{Luo2024}, we acquire a self-consistent partitioning solution by an iterative procedure. Saturation effects are particularly important for us because, given the rapid sinking speed of the continuously accreted metal, the total metal mass in a layer is tiny compared to the silicates (see Sect. \ref{sec:magmaoceantransport}). For a fixed partition coefficient, as the metal mass decreases, the mass concentration of a volatile partitioned into the metal increases.

By choosing whether or not to include prior low pressure and temperature data, \citet{Tagawa2021} actually present two partition coefficient fits that differ substantially. The water partition coefficient fit that excludes previous studies decreases with increased pressure in the 30 GPa to 60 GPa range. To avoid extrapolation, we use the fit which incorporates previous experiments (this is the one shown in Fig. \ref{partitioncoef}). This fit also shares the feature that water instead becomes more siderophile at higher pressures, similar to \citet{Yuan2020}. Despite their different slopes, the two partition coefficient fits from \citet{Tagawa2021} both predict that water is significantly more siderophile throughout the magma ocean than predicted by the fit from \citet{Luo2024}.

\subsubsection{Carbon} \label{sec:carbon}
For carbon, the experiments of \citet{Fischer2020} found a significant pressure dependency, where carbon becomes several orders of magnitude less siderophile at high core-mantle boundary pressures. \citet{Blanchard2022}, on the other hand, do not find a pressure dependence for the carbon partition coefficient and instead fit their data with only a temperature dependence. Recently, a third diamond anvil cell study of carbon was published \citep{Tsutsumi2025} whose proposed carbon partition coefficient is independent of $p$ and $T$ in the experimental range $33$--$56$ GPa and $3630$--$4760$ K. For our melt composition, \citet{Tsutsumi2025} predicts $D_C \approx 100$, consistent with \citet{Blanchard2022} and inconsistent with \citet{Fischer2020}. We therefore choose the carbon partition coefficient fit of \citet{Blanchard2022} to represent our fiducial model.

\subsubsection{Nitrogen} \label{sec:nitrogen}
Similarly to carbon, it is unclear whether the nitrogen partition coefficient depends on pressure, temperature, or both \citep{Roskosz2013, Grewal2019, Speelmanns2019, Jackson2021, Shi2022, Huang2024}. In particular, different studies infer positive \citep{Grewal2019, Jackson2021, Shi2022}, negative \citep[e.g.][]{Roskosz2013}, or negligible \citep[e.g.][]{Speelmanns2019, Huang2024} pressure dependence of the nitrogen partition coefficient.

We choose \citet{Huang2024} to represent our fiducial nitrogen partition coefficient due to the high pressures ($\sim$$100$~GPa) and temperatures ($\sim$$5000$~K) probed. Their proposed partition coefficient fit is fairly lithophile compared to previous work \citep[e.g.][]{Grewal2019, Jackson2021, Shi2022}. To explore a more siderophile alternative, we adopt the nitrogen partition coefficient of \citet{Grewal2019} to represent our non-fiducial case. We note, however, that this requires extrapolating the partition coefficient fit of \citet{Grewal2019} to pressures well above the $<$$6$~GPa tested in their experiments (Fig. \ref{partitioncoef}).

\subsection{Magma ocean transport} \label{sec:magmaoceantransport}
As pebbles and planetesimals accrete to the planet, the fate of their mass components differ. We assume that the volatile component of pebbles and planetesimals first enters the atmosphere, that their silicate component grows the top layer of the magma ocean, and that their metal component sinks through the magma ocean, rapidly reaching the core of the planet \citep[e.g.][]{Rubie2003, Modirrousta2025}. These sedimenting metal droplets transport dissolved volatiles to the core of the planet and the presence of such volatiles in the core may partially explain the inferred underdensity of the core relative to pure Fe or Fe-Ni alloy \citep[e.g.][]{Alfe2007, Badro2007, Hirose2021}. Volatile sedimentation to the core importantly also reduces the volatile inventory for prebiotic and biotic chemistry on the surface. Sedimentation (see Sect. \ref{sec:sedimentation}) transports the volatiles partitioned in the metal, whereas convection (see Sect. \ref{sec:convection}) causes transport of the volatiles partitioned in the silicate melt. 

Small metal droplets generally reach metal-silicate partitioning equilibrium of their embedded volatiles with the surrounding magma after falling less than 200 m \citep{Rubie2003}. Therefore, assuming local equilibrium in a layer $k$, the total mass of a volatile $i$ in the layer, $M_{ik}$, can be expressed with the mass of metal $M_{\text{met},k}$ and silicate $M_{\text{sil},k}$ in the layer, as well as their corresponding local equilibrium mass fractions $C_{ik}^\text{(met)}$ and $C_{ik}^\text{(sil)}$, as
\begin{equation}
    M_{ik}=C_{ik}^\text{(met)} M_{\text{met}, k} + C_{ik}^\text{(sil)} M_{\text{sil}, k}.
\end{equation}
The mass of metal in a given layer is the mass flux of accreted metal multiplied by the residence time in the magma ocean layer,
\begin{equation}
    M_\text{met} = \dot{M}_\text{p} f_\text{met} \frac{\Delta r}{v_\text{met}}, \label{eq:Mmet}
\end{equation}
where $\dot{M}_\text{p}$ is the planet growth rate and $f_\text{met}$ is the metal fraction of the accreted material. We have removed the subscript $k$ because $M_\text{met}$ is constant in all equal-width layers assuming a constant sedimentation speed (see Sect.\,\ref{sec:sedimentation}).
Given the rapid sedimentation speed, $M_\text{met}/M_\text{sil} \approx 10^{-8}$--$10^{-7}$, increasing with depth. Using the definition of the partition coefficient (Eq. \ref{eq:partitioncoef}) we derive expressions for the volatile mass partitioned into the metal and the silicate melt respectively at equilibrium \citep[e.g.][]{Anatomy2},
\begin{align}
    & M_{ik}^ \text{(met)} = M_{ik} \, \frac{D_{ik}M_{\text{met}}}{M_{\text{sil}, k} + D_{ik} M_{\text{met}}}, \label{eq:Mikmet} \\
    & M_{ik}^\text{(sil)} = M_{ik} \, \frac{M_{\text{sil},k}}{M_{\text{sil}, k} + D_{ik} M_{\text{met}}} \label{eq:Miksil}.
\end{align}
Siderophile volatiles with $D_{ik} \gg 1$ preferentially partition into the metal droplets whereas lithophile volatiles with $D_{ik} \ll 1$ preferentially partition into the silicate melt. 

\citet{Anatomy2} showed that the entire mantle inventory of volatiles has a steady-state solution matching the global metal-silicate partitioning equilibrium 
\begin{equation}
    M_i^\text{(mantle)} = M_i^\text{(tot)} \frac{M_\text{mantle}}{M_\text{mantle}+D_i M_\text{core}} \label{eq:globalequilibrium}
\end{equation}
in the limit of continuous volatile delivery, no atmosphere, and constant $D_i$
(Appendix \ref{sec:sedimentationcontinuous}). 
Note the similarity to Eq. \ref{eq:Miksil}. However, thermal processing of pebbles sublimates the volatile component of pebbles after the envelope reaches sufficient temperatures, thereby stopping volatile accretion. In this early volatile delivery case, volatile delivery ceases once the planet reaches a mass $M_{\text{p},0}$ and the envelope temperature is hot enough, fixing the total volatile inventory thereafter (see Sect. \ref{sec:planetgrowth}). Core growth, however, continues and removes a fraction of volatiles from the magma ocean as the metal sediments (see Sect. \ref{sec:sedimentation}). Without continued volatile delivery, the magma ocean reservoir progressively depletes as the planet grows. We derive for this case in Appendix \ref{sec:twothreecomponentderivation} that
\begin{equation}
    M_i^\text{(mantle)} = M_i^\text{(tot)} \left( \frac{M_{\text{p},0}}{M_\text{p}}\right)^{\frac{D_i M_\text{core}}{M_\text{mantle}}},\label{eq:twocomponent}
\end{equation}
using constant $D_i$ and in the limit case where the volatiles fully dissolve in the magma ocean (see Sect. \ref{sec:dissolution}).
Notably, because $D_i$ now enters the exponent of the mass ratio, in contrast to the linear dependence in Eq. \ref{eq:globalequilibrium}, the fate of the volatile is highly sensitive to its partitioning. If the volatile is soluble, siderophile ($D_i \gg 1$) and delivered early $(M_{\text{p},0} \ll M_\text{p})$, then almost all of the volatile mass will enter the core by runaway sedimentation. 

The sensitive partition coefficient dependence of volatile sedimentation in the early delivery scenario is reduced for atmophile volatiles.
In the case that the atmospheric inventory of a volatile species cannot be neglected, we also derive (Appendix \ref{sec:twothreecomponentderivation}) a general solution for the mass of volatile $i$ in the mantle and the atmosphere by assuming dissolution equilibrium between the atmosphere and the magma ocean,
\begin{equation}
    M_i^\text{(mantle)}+M_i^\text{(atm)} = M_i^{\text{(tot)}}\left(\frac{M_{\text{p},0}^{2/3}+\gamma_i}{M_\text{p}^{2/3}+\gamma_i} \right)^{\frac{3}{2}\frac{D_i M_\text{core}}{M_\text{mantle}}} \label{eq:threecomponent}
\end{equation}
where $\gamma_i$ is a coefficient that relates the volatile masses in the atmosphere and in the magma ocean,
\begin{equation}
    M_i^\text{(atm)} = \gamma_i M_\text{p}^{-\frac{2}{3}} M_i^{\text{(mantle)}}, \label{eq:gammarelation}
\end{equation}
and is inversely proportional to the solubility coefficient $\alpha_i$ (see Sect. \ref{sec:dissolution}) and the mean molecular weight $\bar{\mu}$,
\begin{equation}
    \gamma_i = \frac{1}{G\left(1-f_\text{met}\right)\alpha_i} \left(\frac{3^4}{4\pi\bar{\rho}^4}\right)^{1/3} \frac{\mu_i}{\bar{\mu}}, \label{eq:gamma}
\end{equation}
where $\bar{\rho}$ is the average density of the planet. Equations \ref{eq:threecomponent}--\ref{eq:gamma} assumes that the volatile species has a Henrian (linear) solubility.
Equation \ref{eq:threecomponent} also assumes that the mean molecular weight of the atmosphere $\bar{\mu}$ is constant. If a volatile species $i$ is very soluble, then this general solution (Eq. \ref{eq:threecomponent}) reduces to the limit case of full dissolution (Eq. \ref{eq:twocomponent}) with $M_i^\text{(atm)}=0$. In the opposite limit where the volatile is fairly insoluble ($\gamma_i \gg M_\text{p}^{2/3}$) then Eq. \ref{eq:threecomponent} reduces to $M_i^\text{(atm)} \approx M_i^\text{(tot)}$.

In general, Eqs. \ref{eq:twocomponent} and \ref{eq:threecomponent} demonstrate that early delivery of volatiles enhances sedimentation to the core. Furthermore, Eq. \ref{eq:threecomponent} shows that combining dissolution and metal-silicate partitioning causes core sedimentation to depend sensitively on volatile delivery timing, solubilities, and partition coefficients.

\subsubsection{Sedimentation} \label{sec:sedimentation} 
Metal droplets, of diameter $\sim$cm \citep{Rubie2003}, have a much greater surface area-to-volume ratio than the core. We therefore assume that there is no exchange of volatiles between the core and mantle except for sedimentation of volatiles bound in such metal droplets.
We resolve the magma ocean into spherical equal-width layers. The time $\tau$ for a metal droplet to sink through a magma ocean layer of width $\Delta r$ can be calculated using the sinking speed of the metal $v_\text{met} \approx 1 \, \text{m/s}$ \citep[e.g.][]{Rubie2003} yielding
\begin{equation}
    \tau = \frac{\Delta r}{v_\text{met}}.
\end{equation}
By construction, $M_{ik}^ \text{(met)}$ is also the mass of volatile $i$ transported through the lower boundary of the $k^\text{th}$ layer over the time $\tau$. The rate of mass change due to sedimentation can then be expressed as the difference between the mass transported through the upper and lower boundary of the layer,
\begin{align}
    \frac{d M_{ik}}{d t}\bigg\vert_{\text{sedi}} &= \frac{M_{i, k+1}^\text{(met)}}{\tau} - \frac{M_{ik}^\text{(met)}}{\tau}.
\label{eq:massratesedi}
\end{align}
Sedimentation stops after the bottom of the magma ocean crystallizes, preventing further volatile transport to the core. Any volatile delivery after such core segregation is generally referred to as the late veneer \citep[e.g.][]{Walker2009, Day2016}, and likely constitutes $<$$0.1$--$0.5$$~\text{wt}\%$ of the Earth's mass \citep{Chou1978, Calvo2026}.

\subsubsection{Convection \label{sec:convection}}
Planet growth supplies terrestrial magma oceans with significant heat flux from the release of the potential energy of the accreted material. Magma oceans transport this heat primarily through convection in the hard turbulence regime
\citep[e.g.][]{Solomatov2000, Solomatov2007, Neumann2014, Monteux2016}. 

Volatiles that are partitioned into the silicate melt are transported by convective motions. Since partition coefficients depend on local pressure and temperature (see Sect. \ref{sec:partitioncoef}), each magma ocean layer partitions a different fraction of its volatile mass between metal droplets and the silicate melt (Eqs.~\ref{eq:Mikmet},~\ref{eq:Miksil}). In principle, this implies that the fraction of volatiles transported by convection versus sedimentation varies with depth in the magma ocean, which could produce volatile concentration heterogeneity. However, the magma mass in a layer greatly exceeds the metal mass, such that essentially all of the volatile mass is carried by the silicate melt -- and hence by convective motion -- regardless of the partition coefficient (Eq.~\ref{eq:Miksil}).

In the hard turbulence regime, convection rapidly mixes heterogeneities in the magma ocean, on timescales that can be as short as minutes \citep{Thomas2024}. We therefore assume that the volatiles are well mixed and that their concentration is uniform throughout the magma ocean.

\subsection{Oxygen fugacity} \label{sec:oxygenfugacity}
In formation scenarios with continuous mass accretion $\dot{M}_\text{acc}$, for example pebble accretion \citep{Johansen2021} or incremental planetesimal accretion \citep[e.g.][]{Rubie2015, Woo2022}, there is a continuous supply of metal (Fe and Ni) to the magma ocean. If the delivery of metallic iron was more sporadic, for example as a result of giant impacts, then oxygen fugacity could vary throughout the magma ocean due to draining of the metal into the core and subsequent oxidation of Fe$^{2+}$ to Fe$^{3+}$ \citep[e.g.][]{Hirschmann2012, Deng2020, Zhang2024ferric}. Since metallic iron is available everywhere in the magma ocean during accretion when considering continuous mass accretion of smaller bodies, the oxygen fugacity of even the upper layers of the magma ocean are buffered by the iron-wüstite redox reaction
\begin{equation}
    \text{Fe}\,\text{(l)} + \frac{1}{2} \, \text{O}_2 \, \text{(g)} \rightleftharpoons \text{FeO} \, \text{(l)} \label{eq:IWreac}
\end{equation}
rather than by
\begin{equation}
    \text{FeO} \, \text{(l)} + \frac{1}{4} \, \text{O}_2 \, \text{(g)} \rightleftharpoons \text{FeO}_{1.5} \, \text{(l)}. \label{eq:FeOtoFeO1.5}
\end{equation}
Equation \ref{eq:FeOtoFeO1.5} would buffer the oxygen fugacity if metallic iron was absent from the magma ocean \citep[e.g.][]{Sossi2020, Deng2020, Ortenzi2020}, which will be the case after the end of the accretion phase. We assume here a constant \citep[e.g.][]{Fischer2020, Huang2021, Anatomy3} log distance, $\Delta \text{IW}=-2$, from a reference oxygen fugacity produced by the iron-wüstite buffer (Eq. \ref{eq:IWreac}; \citealt{Hirschmann2021}), throughout the magma ocean. Importantly, this oxygen fugacity is also imposed on the atmosphere. The equilibrium condition for Eq. \ref{eq:IWreac} can then be used to express the oxygen fugacity,
\begin{equation}
    \log_{10} f\text{O}_2 (\Delta \text{IW}) = \Delta\text{IW}- 2 \log_{10} K_{\ref{eq:IWreac}}, 
\end{equation}
where the iron-wüstite equilibrium coefficient $K_{\ref{eq:IWreac}}$ is known given the pressure and the temperature \citep{Hirschmann2021}. We then define $\Delta \text{IW}$ as
\begin{equation}
    \Delta \text{IW} = 2 \log_{10} \left(\frac{a_\text{FeO}^\text{(sil)}}{a_\text{Fe}^\text{(met)}} \right).
\end{equation}
Here $a_\text{FeO}^\text{(sil)}$ and $a_\text{Fe}^\text{(met)}$ are the activities of FeO in the silicates and Fe in the metal, respectively \citep[e.g.][]{Armstrong2019, Hirschmann2021}. 

\subsection{Atmospheric speciation} \label{sec:speciation}
Volatile gases in the atmosphere speciate according to oxygen fugacity \citep[e.g.][]{Iacono2012, Ortenzi2020},
\begin{equation}
    \text{H}_2(\text{g}) + \frac{1}{2} \text{O}_2(\text{g})  \underset{K_{\text{H}_2}}{\rightleftharpoons} \wat(\text{g}), \label{rec:H2spec}
\end{equation}
\begin{equation}
    \text{CO}(\text{g}) + \frac{1}{2} \text{O}_2(\text{g}) \underset{K_{\text{CO}}}{\rightleftharpoons} \car(\text{g}), \label{rec:COspec}
\end{equation}
such that \begin{equation}
    p_{i}^{\text{(red)}} \propto 10^{-\Delta \text{IW}/2} \, p_{i}^{\text{(ox)}}, \label{eq:deltaIWspeciation}
\end{equation}
where $p_{i}^{\text{(red/ox)}}$ represents the atmospheric partial pressure of the reduced and oxidized speciation of volatile $i$, respectively, in Eqs. \ref{rec:H2spec} \ref{rec:COspec}. High oxygen fugacity results in more oxidized volatile species ($\wat, \car$) and less reduced volatile species ($\text{H}_2,$ CO). Nitrogen speciates as $\nit$ at high temperatures, irrespective of oxygen fugacity. We deliberately excluded any speciation as $\text{CH}_4$ and NH$_3$, since only negligible amounts exist at the surface temperature of the magma ocean \citep[e.g.][]{Ortenzi2020}. Neglecting CH$_4$ and NH$_3$ also allows us to study volatile exchange between planet reservoirs in parallel without coupled interactions between volatiles.

\subsection{Dissolution} \label{sec:dissolution}
Since the surface temperature remains below the mantle critical temperature of $\sim$$7000$ K \citep{Caracas2023}, there exists a well-defined boundary between the liquid magma ocean and the gaseous atmosphere.
The top of the magma ocean and the bottom of the atmosphere interact by dissolving gas in the magma melt and outgassing from the melt into the atmosphere \citep[e.g.][]{Lichtenberg2021}. Some volatile-bearing gaseous species are more soluble than others and therefore more easily incorporated into the magma melt. We solve the dissolution equilibrium of the volatiles by exchanging the volatile mass between the atmosphere and the top layer of the magma ocean while conserving the total volatile mass in these reservoirs. 

\begin{table}[]
\caption{Parameters for the magma-atmosphere solubility of volatile elements described by Eq.\, \ref{eq:Licht}. }
\label{tab:solubility} 
\centering
\begin{tabular}{lll}
\rowcolor[HTML]{CBCEFB}
\bottomrule Volatile & $\alpha$ [ppmw Pa$^{-1/\beta}$] & $\beta$ \\ \hline
$\wat$  & $1.033 \times 10^0$ & 1.747 \\
$\car$  & $4.400 \times 10^{-6}$ & 1     \\
$\nit$  & $6.110 \times 10^{-7}$ & 1     \\
H$_2$   & $2.315 \times 10^{-5}$ & 1     \\
CO      & $5.500 \times 10^{-7}$ & 1     
\end{tabular}
\tablefoot{Solubility parameters modified from Table 1 in \citet{Lichtenberg2021}. We use different sources for the solubility of the following gas species $\car$: \citep{Papale1997, Hirschmann2012, Anatomy2, Mysen1975, Stolper1988, Pan1991, Blank1993, Dixon1995}; $\nit$: \citep{Sossi2020, Anatomy2, Bernadou2021}; CO: \citep{Armstrong2015, Hirschmann2016}.
}
\end{table}

We model the dissolution of all volatiles species using solubility fits of the form
\begin{equation}
    X_i^{(\text{melt})} = \alpha_i \, p_i^{1/\beta_i}, \label{eq:Licht}
\end{equation}
where $X_i^{(\text{melt})}$ is the mass fraction of volatile $i$ dissolved in the top layer of the magma ocean, $\alpha_i$ is the dissolution coefficient and $\beta_i$ is analogous to the stoichiometry coefficient of the dissolution reaction \citep[e.g.][]{Lichtenberg2021}. For example, assuming $\wat$ dissolves purely as 2 OH$^{-}$ yields $\beta_{\wat} = 2$. In actual experiments \citep[e.g.][]{Silver1990, Holtz1995, Moore1998, Gardner1999, Yamashita1999, Liu2005}, where part of the $\wat \, (\text{g})$ dissolves as $\wat \, \text{(melt)}$ \citep[e.g.][]{Stolper1982, Burnham1974, Burnham1994}, a value of $\beta_{\wat} = 1.747$ is found instead \citep[][]{Lichtenberg2021}. 

We dissolve $\nit$ according to a simplified Henrian ($\beta_{\nit} = 1$) solubility \citep[][]{Sossi2020, Anatomy2}. For our redox conditions ($\Delta\text{IW}>-3$), nitrogen dissolution is effectively linear and $f\text{O}_2$-independent \citep[e.g.][]{Bernadou2021}. Similarly, we also model the solubility of $\car$ as Henrian. $\car$ dissolves close to Henry's law at $p_{\car}< 1000$ bar \citep[e.g.][]{Mysen1975, Stolper1988, Pan1991, Blank1993, Dixon1995, Papale1997, Hirschmann2012, Anatomy2, Lichtenberg2021}. 
We use the Henrian carbon solubility of \citet{Armstrong2015} under reducing conditions ($\Delta \text{IW} = -2$) following the methods of \citet{Hirschmann2016} and \citet{Maurice2024}. We adopt the H$_2$-solubility fit from experiments at $< \,$$3$ GPa and 1400 K \citep{Hirschmann2012solubility}. A recent study of solubility at sub-Neptune magma ocean conditions ($16$--$60$ GPa, $4000$--$4700$ K) found a strong temperature dependence and a weak pressure dependence for H$_2$ \citep{Miozzi2025}. Their experimental fit would suggest nearly three times higher solubility of H$_2$ than \citet{Hirschmann2012solubility} at our terrestrial magma ocean surface temperature of 3000 K (see Fig. \ref{widestructure}). However, using this fit would require extrapolating both to lower temperatures and much lower pressures than the experimental conditions of \citet{Miozzi2025}. Instead, we adopt the H$_2$ solubility fit of \citet{Hirschmann2012} but note that H$_2$ may, in fact, be more soluble. Table \ref{tab:solubility} lists the solubility parameters used for volatile dissolution.

\section{Results} \label{sec:results} 
We test both early volatile delivery (representing the pebble accretion end member case) and continuous volatile delivery (representing the planetesimal accretion end member case). 
The outcomes of volatile transport are sensitive to the choice of partition coefficients, given the large discrepancies between \citet{Tagawa2021} and \citet{Luo2024} for $\wat$, between \citet{Fischer2020} and \citet{Blanchard2022} for $\car$, and between \citet{Grewal2019} and \citet{Huang2024} for $\nit$. 
For our fiducial model (see Sect. \ref{sec:partitioncoef}), we use the partition coefficients of \citet{Luo2024, Blanchard2022, Huang2024}. Although we focus our presentation on the fiducial model (the L-B-H model), we present selected results from the alternative model (the T-F-G model), which uses the partition coefficients of \citet{Tagawa2021, Fischer2020, Grewal2019}. Given Earth's uncertain core hydrogen content ($9$ -- $45$ Earth oceans), we also vary between low (Lo) and high (Hi) hydrogen budget for our proto-Earth models (See Sect. \ref{sec:planetgrowth}). We slightly focus on the results of the Lo models given closer alignment to previous estimates of Earth's hydrogen content, but also present results from the Hi models to highlight the volatile inventories' dependence on the total volatile budget. 

In Sect. \ref{sec:fate} we detail the volatile inventories in each of the reservoirs of the planet at the end of accretion and how they change as the planet grows. 
In Sect. \ref{sec:effective} we quantify the combined effect of solubility and delivery timing on how volatiles partition into the core by constructing effective partition coefficients.

\subsection{Fate of the volatiles} \label{sec:fate}
\begin{figure*}[!t]
\centering
\includegraphics[width=\textwidth]{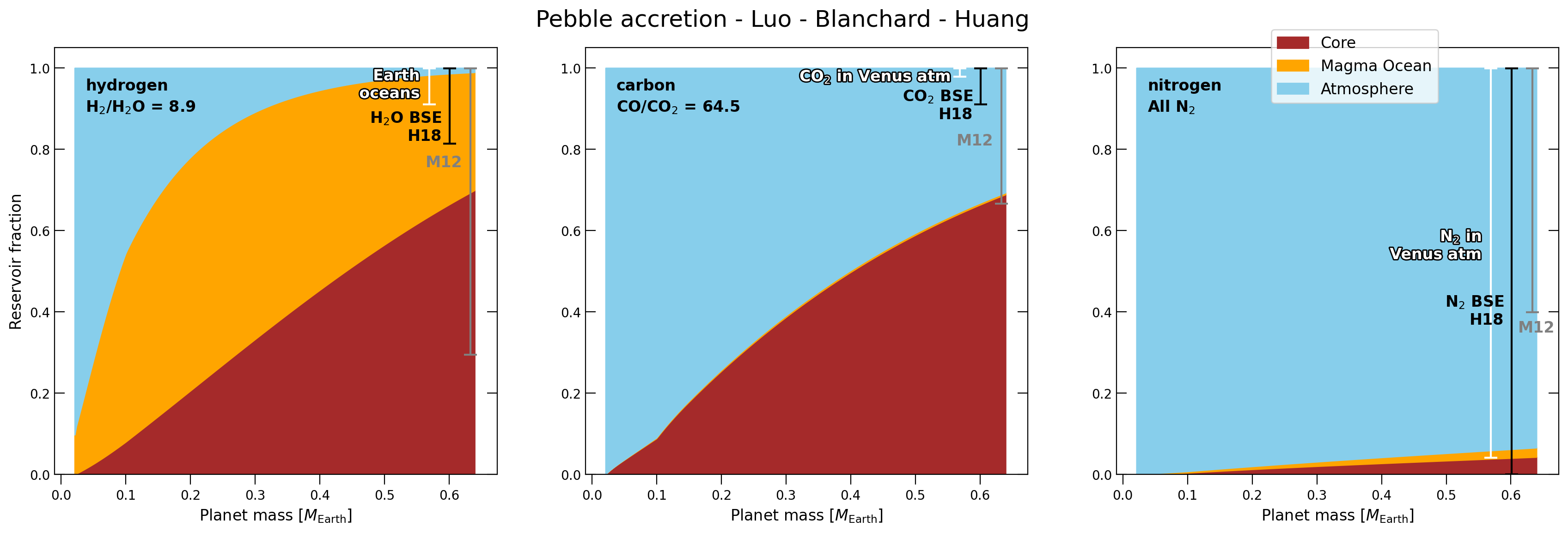}
\includegraphics[width=\textwidth]{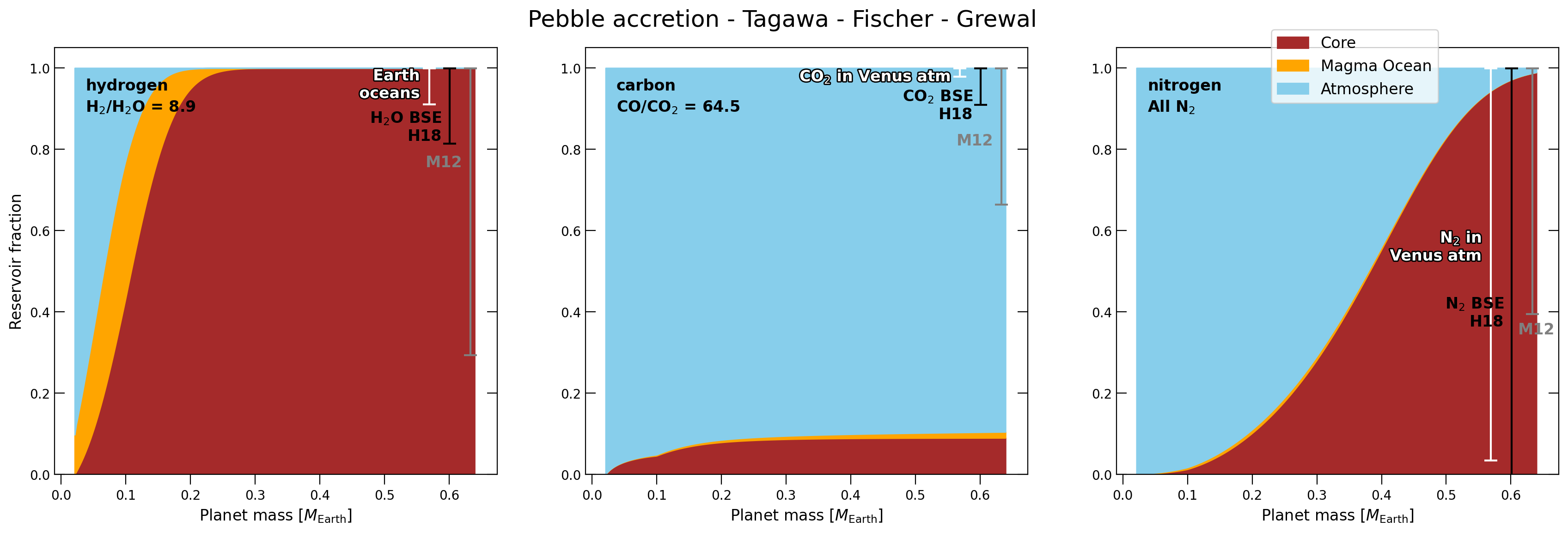}
\caption{Volatile inventory for our pebble accretion model in the core (red), magma ocean (orange) and atmosphere (blue) as the planet grows from $0.02 \ M_\text{E}$ to $0.64 \ M_\text{E}$ using the fiducial model partition coefficient fits of Luo - Blanchard - Huang (top row) or the partition coefficient fits of Tagawa - Fischer - Grewal (bottom row). The height of each color represents, for a given instantaneous planet mass, the mass of the volatile that is in each reservoir normalized to the global mass of the volatile. Volatile masses are reported as the mass of the oxidized species of the volatile ($\wat$, $\car$, $\nit$). The molar ratios between reduced and oxidized volatile species, set by the oxygen fugacity (Eq.~\ref{eq:deltaIWspeciation}), are shown in the top left of each panel. Volatile reservoirs of Venus and Earth (white text and bar) are included for comparison. BSE ranges indicate the respective volatile masses in bulk silicate Earth (black text and bar, \citealt{Hirschmann2018}; gray text and bar \citealt{Marty2012}). The fiducial model (top row) lies between estimates of the BSE content of $\wat$ and C, and matches one estimate of the BSE content of N.}
\label{widereservoirpebble}
\end{figure*}

\begin{figure*}[]
\centering
\includegraphics[width=\textwidth]{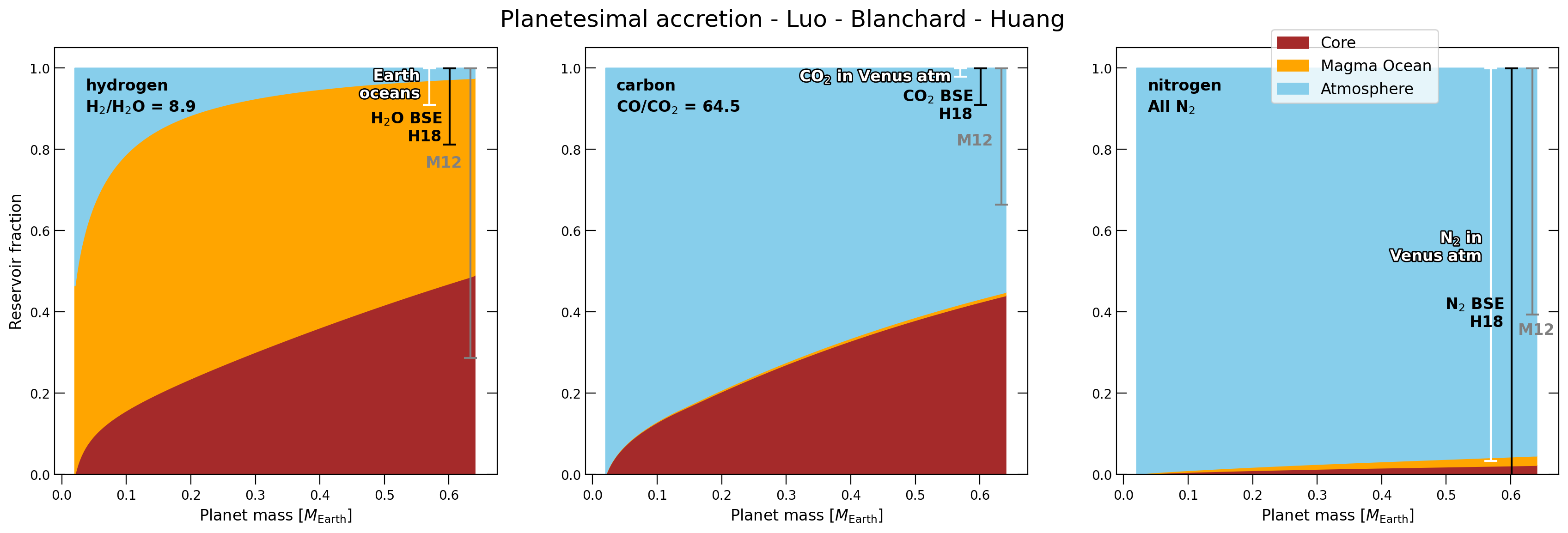}
\includegraphics[width=\textwidth]{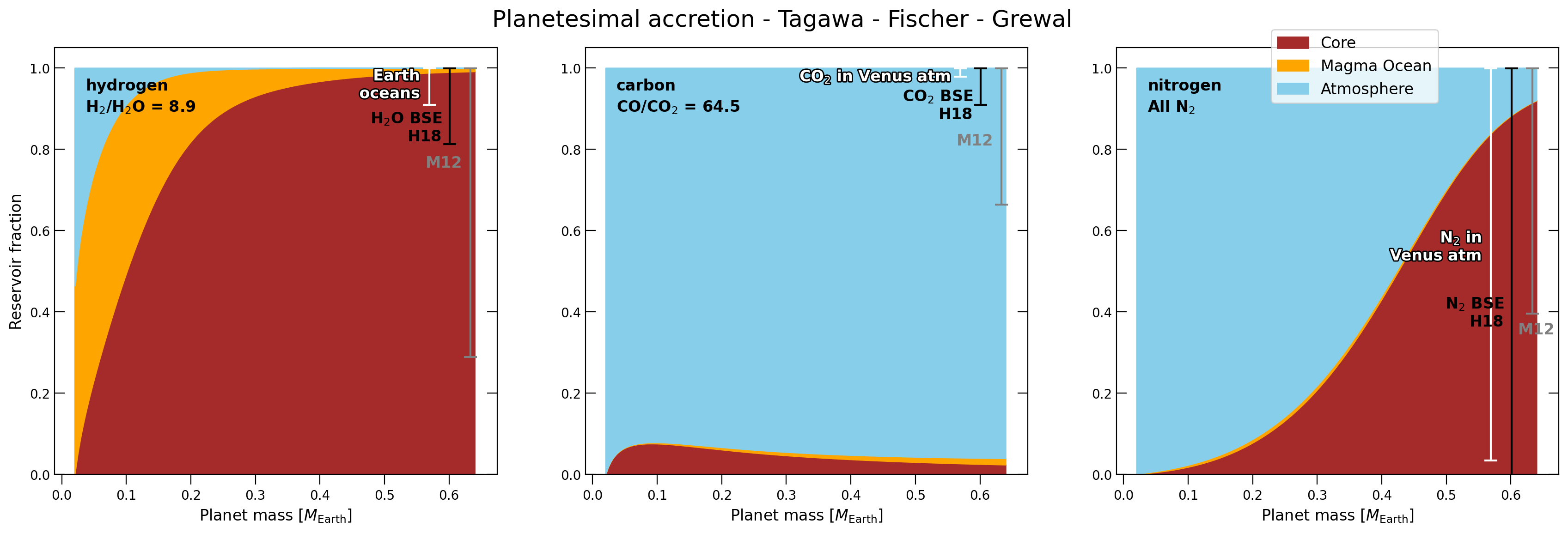}
\caption{Same as Fig. \ref{widereservoirpebble} but for planet grown by planetesimal accretion with continuous volatile delivery (see Sect. \ref{sec:planetgrowth}). Both the L-B-H model (top row) and the T-F-G model (bottom row) produce higher atmosphere and magma ocean contents for $\wat$, C, and N than the corresponding pebble accretion scenario (Fig. \ref{widereservoirpebble}) due to the effect of later volatile delivery (see Sect. \ref{sec:magmaoceantransport}). For carbon using the \citet{Fischer2020} partition coefficient in particular, sedimentation to the core also reduces in the planetesimal case because significant carbon accretes after becoming lithophile at the increasing pressure of the core-mantle boundary.}
\label{widereservoirplanetesimal}
\end{figure*}

In this section, we show the volatile inventories of the planetary reservoirs resulting from our models of proto-Earth.
Figures \ref{widereservoirpebble} and \ref{widereservoirplanetesimal} show how each volatile element distributes between the three planet reservoirs, as the planet grows in our pebble and planetesimal accretion models with early volatile delivery and continuous volatile delivery, respectively (see Sect. \ref{sec:planetgrowth}). 
Table \ref{tab:BSE} and Fig. \ref{bubbleBSE} lists the final volatile inventories of the bulk silicate reservoir of proto-Earth resulting from models that vary its formation pathway, total hydrogen budget, adopted set of partition coefficient fits, volatile solubilities, and oxygen fugacity. Figures \ref{piatm} and \ref{masschangeboth} show the evolution of the outgassed atmosphere, resulting from ongoing volatile sedimentation to the core. 

\subsubsection{Reservoir inventories, pebble accretion model}
The top row of Fig. \ref{widereservoirpebble} shows the pebble accretion model using the L-B-H partition coefficients. Hydrogen dissolves easily into the magma ocean, causing the atmospheric hydrogen to gradually drain into the magma ocean as the planet increases its mass and hence its capability to store volatiles in the magma. The hydrogen then partitions into the sinking metal and sediments to the core (see Sect. \ref{sec:sedimentation}). At the end of the protoplanetary disk phase, before any putative giant impacts, $\sim$$70\%$ of the hydrogen mass is in the core and the remainder is almost entirely in the magma ocean. The total hydrogen mass contained in the magma ocean and atmosphere at the end of the simulation corresponds to $4.6 \times 10^{21}$ kg of $\wat$, equivalent to approximately 3.4 times the mass of Earth's oceans. Unlike $\wat$, carbon is atmophile and the dissolution equilibrium favors outgassing irrespective of the protoplanet's mass. However, carbon is significantly more siderophile than hydrogen ($D_\text{C} \gtrsim 100$; Fig. \ref{partitioncoef}) resulting in $\sim$$69\%$ core sequestration. Carbon sediments into the core efficiently despite the low carbon mass present in the magma ocean at any point in time. Any carbon dissolved from the atmosphere into the magma ocean quickly sediments into the core. The remaining bulk silicate carbon mass at the end of the pre-giant impact accretion is mostly atmospheric CO, equivalent to $7.3 \times 10^{21}$ kg of $\car$ or roughly 15.5 times the $\car$ mass of Venus' atmosphere. This is significantly higher than the mass of carbon-bearing species in the atmosphere in the models of \citet{Anatomy3}. We attribute this difference in results to our model having relatively insoluble CO instead of $\car$ as the main carbon-bearing atmospheric species. Nitrogen actually has a higher partition coefficient than hydrogen for all $(p, T)$ (see Sect. \ref{sec:partitioncoef}), yet only $\sim$$4$$\%$ of the nitrogen enters the core. This is because water is much more soluble than nitrogen (Table \ref{tab:solubility}) and therefore has a higher mass fraction in the magma ocean, which increases the sedimentation efficiency \citep[e.g.][]{Speelmanns2019, Anatomy3}. In the end, only $\sim$$2 \%$ of the nitrogen is in the mantle and the remaining $\sim$$93\%$ of the nitrogen is in the atmosphere, $M_{\nit \text{(g)}} \approx 1.1 \times 10^{19}$ kg, equal to the mass of nitrogen in Venus' atmosphere \citep{Way2016}.

The bottom row of Fig. \ref{widereservoirpebble} shows how volatiles distribute between the planet reservoirs for the pebble accretion model using the T-F-G partition coefficients. Due to the much more siderophile affinity of hydrogen in this case, compared to the \citet{Luo2024} partition coefficient, hydrogen rapidly enters the core and already after $M_\text{p} \approx 0.24 \, M_\text{E}$ less than 1\% of the hydrogen mass remains outside the core. Although hydrogen continuously sediments from the magma ocean to the core for both choices of partition coefficient, the depletion is extreme in this more siderophile scenario. The extreme sensitivity to $D_{\wat}$ is because hydrogen is delivered early in the pebble accretion model and dissolves well into the magma. Therefore, we approach the regime where Eq. \ref{eq:twocomponent} applies and essentially all of the hydrogen enters the core. 

For carbon, \citet{Fischer2020} predicts less siderophile affinity than \citet{Blanchard2022}, implying less carbon sedimentation into the core. 
As the planet grows, carbon struggles to sediment further as it becomes lithophile near the core; $D_\text{C, CMB} \lesssim 1$
after exceeding $M_\text{p} \approx 0.3 \, M_\text{E}$ and $p_\text{CMB} \gtrsim 40$ GPa, $T_\text{CMB} \gtrsim 4500$ K.  The slowing rate of carbon sedimentation also allows the magma ocean inventory of carbon to hold a few percent of the total carbon mass. At the end of accretion in the \citet{Fischer2020} case, 90\% of the carbon mass is in the atmosphere, corresponding to roughly 2150 bar of CO. The inability of carbon to sediment after the planet reaches a certain $p_\text{CMB}$ increases the dependence on the timing of volatile delivery in the \citet{Fischer2020} case. 

Unlike the L-B-H model, in which nitrogen resides primarily in the atmosphere, the nitrogen partition coefficient of \citet{Grewal2019} predicts a much stronger siderophilic affinity, leading to 99\% of the nitrogen sedimenting into the core. The remaining nitrogen is mostly in the atmosphere.

\subsubsection{Reservoir inventories, planetesimal accretion model}
Figure \ref{widereservoirplanetesimal} shows the volatile contents of the planetary reservoirs of our proto-Earth analogues grown by planetesimal accretion. In contrast to the pebble accretion model (Fig. \ref{widereservoirpebble}), the corresponding planetesimal accretion models produce lower volatile contents in the core and therefore higher volatile contents in both the magma ocean and the atmosphere. This is because, in the planetesimal accretion case, continuous volatile delivery means that the magma ocean contains less volatile mass during early core growth, reducing sedimentation to the core (Sect. \ref{sec:magmaoceantransport}). The top row of Fig. \ref{widereservoirplanetesimal} shows the results of the model using the L-B-H partition coefficients. At the end of the simulation, the magma ocean and atmosphere contain hydrogen equivalent to $7.8 \times 10^{21}$~kg of $\wat$, corresponding to $\sim\,$$5.7$ Earth oceans and nearly twice the hydrogen content in the magma ocean and atmosphere of the pebble-grown proto-Earth model with identical partition coefficients (Fig. \ref{widereservoirpebble}, top row). Similarly, the magma ocean and atmosphere of the planetesimal accretion model also contain nearly twice the carbon content of the corresponding pebble accretion model, equivalent to $1.3\times 10^{22}$ kg of $\car$ or roughly 28 times the $\car$ mass of Venus' atmosphere. Unlike hydrogen and carbon, the atmosphere contains the vast majority of the nitrogen mass in both the pebble accretion and planetesimal accretion scenarios due to its low solubility and its only moderately siderophilic affinity in the L-B-H model. 

The bottom row of Fig. \ref{widereservoirplanetesimal} shows the volatile contents of the planetary reseroirs of the planetesimal accretion model using the T-F-G partition coefficients. As in the corresponding pebble accretion model (Fig. \ref{widereservoirpebble}, top row), effective dissolution combined with the highly siderophile partition coefficient of water described by \citet{Tagawa2021} results in most of the hydrogen entering the core through sedimentation. However, the later delivery timing of volatiles in the planetesimal accretion model allows the magma ocean to retain a small fraction of water in the magma ocean at the end of accretion, equivalent to 0.1 Earth oceans. Carbon behaves similarly in both the pebble accretion and planetesimal accretion scenario using the T-F-G partition coefficients: limited sedimentation occurs at higher planet masses since carbon becomes lithophile at higher CMB pressures. In the planetesimal model (Fig. \ref{widereservoirplanetesimal}, bottom row), the fraction of carbon in the core therefore decreases over time because the total mass of carbon increases with continuous volatile delivery.

\subsubsection{Bulk silicate inventory}

The fate of volatiles can also be represented with a single measure. The bulk silicate concentration of a volatile is defined as the total mass of the volatile in both the atmosphere and the mantle divided by the mass of the mantle. Therefore, the bulk silicate concentration measures the mass of volatiles that did not sediment into the core. The "Early" row of Table \ref{tab:BSE} compares estimates of the bulk silicate Earth (BSE) concentrations of volatile elements \citep{Marty2012, Hirschmann2018, Marty2020} with the corresponding concentrations for our proto-Earth analogue grown by pebble accretion and using both sets of partition coefficient sources for $\wat$ \citep{Tagawa2021, Luo2024}, C \citep{Fischer2020, Blanchard2022} and N \citep{Grewal2019, Huang2024}. Our model ignores volatile delivery by the Moon-forming giant impact, instead accreting the full volatile inventory of Earth during the growth of proto-Earth (see Sect. \ref{sec:planetgrowth}). To then enable comparison to a fully grown Earth, we calculate our bulk silicate comparison values by dividing our volatile masses by the mantle mass of present day Earth (see Sect. \ref{sec:theia} for discussion). 
The bulk silicate concentration of hydrogen in our Lo: L-B-H pebble accretion model lies between the BSE concentrations estimated by \citet[][]{Marty2012, Hirschmann2018}, but outside of either of their reported uncertainties. The carbon concentration lies within the estimated BSE range of \citet{Marty2012} and \citet{Marty2020} but above the estimate of \citet{Hirschmann2018}. 
The nitrogen BSE concentrations estimated by \citet{Hirschmann2018, Marty2020} overlap the results of our Lo: L-B-H pebble accretion model.  

\begin{table}[]
\caption{Volatile abundances in the bulk silicate Earth (BSE) compared to proto-Earth analogues growing from early (pebble) or continuous (planetesimal) accretion of volatiles.}
\label{tab:BSE} 
\centering
\begin{tabular}{lllll}
\rowcolor[HTML]{CBCEFB} \bottomrule
& BSE [ppmw] & $\wat$ & C & N
\\ \hline
\multirow{3}{*}{\rotatebox{90}{Earth\hspace{0.25em}}} & H18\tablefootmark{a} & $710\pm90$ & $140\pm40$ & $2.8\pm0.6$ \\ 
& M12\tablefootmark{b} & $2699\pm1347$ & $526\pm206$ & $1.68\pm0.85$ \\ 
& M20\tablefootmark{c} & & 337, 479 & $2.65 \pm 0.55$ \\ \toprule
\multirow{4}{*}{\rotatebox{90}{Early\hspace{0.25em}}} 
& Lo: L-B-H\tablefootmark{d} & 1157 & 495 & 2.7 \\
& Hi: L-B-H\tablefootmark{d} & 9035 & 967 & 2.8 \\ 
& Lo: T-F-G\tablefootmark{e} & 0 & 1446 & 0 \\ 
& Hi: T-F-G\tablefootmark{e} & 0 & 1574 & 0  \\ \hline 
\multirow{4}{*}{\rotatebox{90}{Conti. \hspace{0.25em}}} 
& Lo: L-B-H\tablefootmark{d} & 1948 & 881  & 2.7 \\ 
& Hi: L-B-H\tablefootmark{d} & 10622 & 1138 & 2.7 \\
& Lo: T-F-G\tablefootmark{e} & 31   & 1537 & 0.2 \\ 
& Hi: T-F-G\tablefootmark{e} & 136 &  1527 & 0.2 \\ 
\bottomrule 
\end{tabular}
\tablefoot{Volatile BSE concentrations are computed as the volatile element masses in the mantle and atmosphere normalized to the mass of the silicate mantle of Earth. Models have either early or continuous accretion of volatiles representing pebble accretion and planetesimal accretion respectively. The early models accrete all their volatiles before the planet mass exceeds mass thresholds for pebble sublimation in the envelope of the planet whereas the continuous models accrete their volatiles throughout their entire growth (see Sect. \ref{sec:planetgrowth}). Total water content of proto-Earth is modeled by assuming that Earth's core contains hydrogen equivalent to either 9 (Lo) or 45 (Hi) Earth oceans \citep[][]{Huang2026}. Section \ref{sec:partitioncoef} describes the choice of partition coefficient parametrizations (either L-B-H or T-F-G). \tablefoottext{a}{Values retrieved from \citep{Hirschmann2018}.} \tablefoottext{b}{Values retrieved from \citep{Marty2012}.} \tablefoottext{c}{Values retrieved from \citep{Marty2020}, providing two estimates of BSE carbon with quartile ranges $\left( 337^{\text{Q1:}211}_{\text{Q3:}504}, 479^{\text{Q1:}323}_{\text{Q3:}646} \right)$ based on assuming the median or the mean C/N ratio of mantle-derived $\car$ gases.} \tablefoottext{d}{Results from fiducial model using the $\wat$, C and N metal-silicate partition coefficients from \citet{Luo2024}, \citet{Blanchard2022} and \citet{Huang2024} respectively.} \tablefoottext{e}{Results from model using the $\wat$, C and N metal-silicate partition coefficients from \citet{Tagawa2021}, \citet{Fischer2020} and \citet{Grewal2019} respectively.}}
\end{table}

For the corresponding high hydrogen budget model -- the Hi: L-B-H pebble model -- the foremost difference is the substantially higher water inventory of the mantle and atmosphere, with the bulk silicate concentration of $\wat$ approaching 1\%. However, the carbon BSE concentration is also significantly higher because the high hydrogen budget produces an atmosphere with lower mean molecular weight $\bar{\mu}^{-1}$ that inhibits dissolution ($M_i^\text{(atm)} \propto \bar{\mu}^{-1}$) (see Eqs. \ref{eq:gammarelation} and  \ref{eq:gamma}). 

Runaway hydrogen sedimentation occurs in the T-F-G pebble accretion models. All of the hydrogen enters the core because the $\wat$ partition coefficient of \citet{Tagawa2021} is highly siderophile, which, coupled with early delivery, causes runaway sedimentation for highly soluble volatiles (See Sect. \ref{sec:magmaoceantransport}). The carbon partition coefficient of \citet{Fischer2020}, which becomes comparatively lithophile at higher pressures, reduces carbon sedimentation to the core and therefore increases the extant carbon in the magma ocean and atmosphere of our proto-Earth analogue at the end of accretion. The bulk silicate concentration of carbon in the Lo: T-F-G model substantially exceeds all of our adopted BSE estimates \citep{Marty2012, Hirschmann2018, Marty2020}. Nitrogen experiences runaway sedimentation to the core in the T-F-G pebble accretion models because the partition coefficient fit of \citet{Grewal2019} is substantially more siderophile than that of \citet{Huang2024} (Fig. \ref{partitioncoef}). 

Table \ref{tab:BSE} also compares the bulk silicate concentrations that result from our proto-Earth models that assume continuous volatile delivery (the "Conti." row), representing our planetesimal accretion model. 
The concentrations in the bulk silicate are consistently larger in the continuous delivery model. Continuous delivery of volatiles produces higher BSE concentrations because volatiles accrete after a significant portion of the metals have already sedimented to the core (see Sect. \ref{sec:magmaoceantransport}). Late-accreted volatiles have less opportunity to sediment and, therefore, remain mostly in the mantle and atmosphere. The BSE concentrations of $\wat$ and C estimated by \citet{Hirschmann2018} do not overlap the results of either the Lo or Hi planetesimal L-B-H model. However, the Lo: L-B-H planetesimal model is consistent with the BSE concentration estimate of \citet{Marty2012} for $\wat$. The model's carbon concentration is slightly above the BSE ranges of \citet{Marty2012, Marty2020}. 

\subsubsection{Dependence on solubility and oxygen fugacity}
\begin{figure}[h!]
\resizebox{\hsize}{!}{\includegraphics{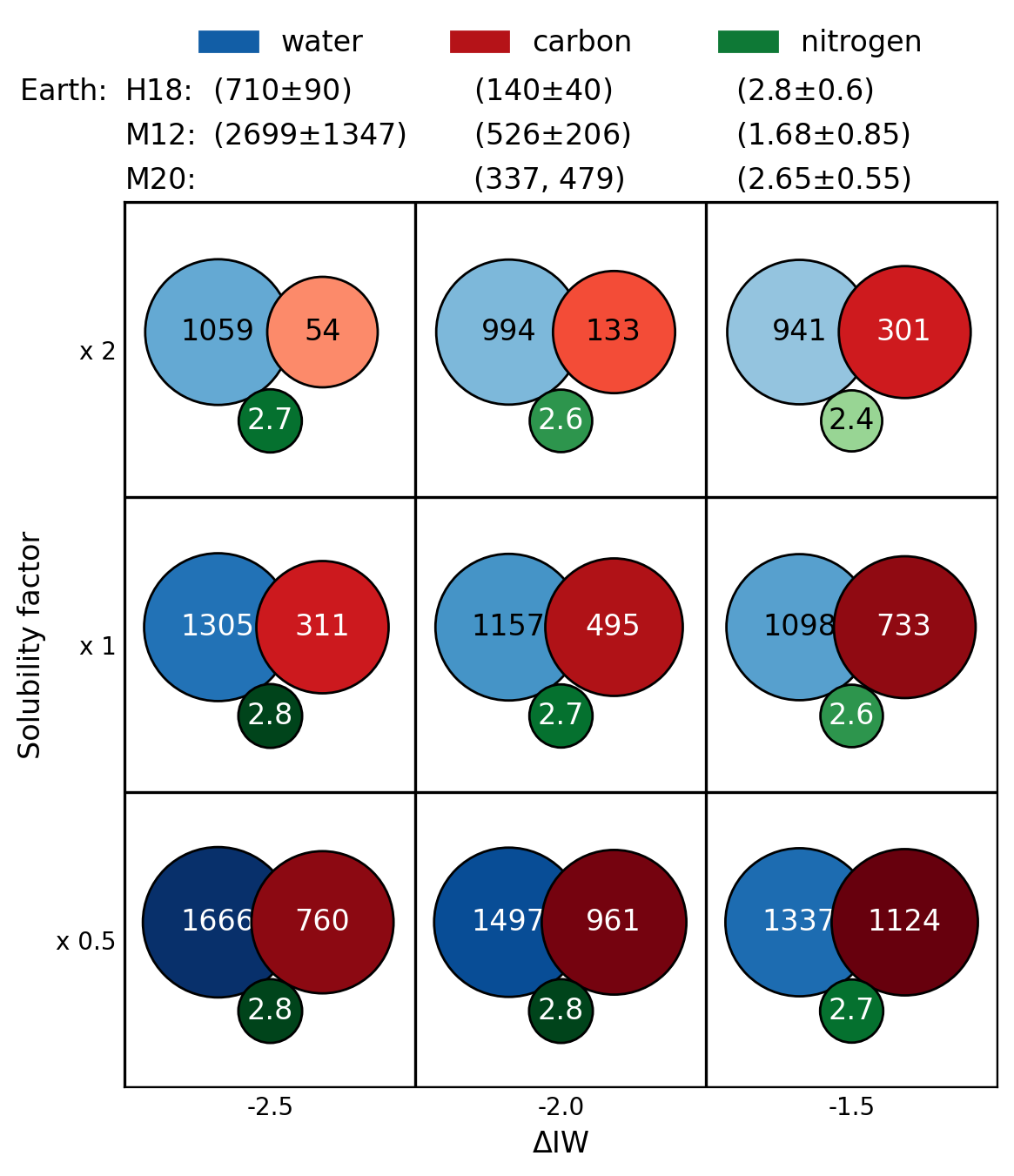}}
\caption{Volatile concentrations in the bulk silicate reservoirs resulting from varying the volatile solubility and oxygen fugacity ($\Delta\text{IW}$; Eq. \ref{eq:IWreac}) of our L-B-H pebble accretion proto-Earth model. The solubility (Eq. \ref{eq:Licht}) of all volatiles were varied by multiplying with the arbitrary factor listed on the y-axis. Both the color intensity and the radius of the circles scale logarithmically with the magnitude of the BSE concentration resulting from the specific model. The central panel represents the fiducial Lo: L-B-H pebble accretion model. 
The reference estimates for bulk silicate Earth volatile concentrations are listed below the title \citep{Hirschmann2018, Marty2012, Marty2020}.
}
\label{bubbleBSE}
\end{figure}

The choice of volatile partition coefficient and whether volatiles are delivered early or continuously affect the resulting BSE concentrations (see Table \ref{tab:BSE}), but the solubility and oxygen fugacity of the magma ocean also impact the results. 
Figure \ref{bubbleBSE} shows the BSE concentrations calculated from Lo: L-B-H pebble accretion models that varied the assumed oxygen fugacity and arbitrarily adjusted the volatile solubility to mimic experimental uncertainty. The bulk silicate concentrations of the volatiles each exhibit different sensitivities to solubility and oxygen fugacity. For all volatiles, BSE concentrations decrease with increasing solubility due to enhanced sedimentation (see Sect.~\ref{sec:magmaoceantransport}). This sensitivity increases for strongly siderophile volatiles, such as carbon (see Eq.~\ref{eq:threecomponent}). Consequently, the model carbon BSE concentrations become nearly four times smaller when the solubility is doubled (top middle panel compared to central panel). The carbon partition coefficient also depends directly on $\Delta$$\text{IW}$ \citep{Blanchard2022}, such that carbon becomes more siderophile under reducing conditions (decreasing $\Delta\text{IW}$). In contrast, the partition coefficient of water does not vary with redox state \citep{Luo2024}. Instead, oxidizing conditions promote hydrogen speciation as $\wat$, which is more soluble than H$_2$. Increasing $\Delta\text{IW}$ therefore enhances hydrogen sedimentation to the core and lowers the hydrogen BSE concentration.
Nitrogen appears insensitive to variations in solubility and oxygen fugacity because its low solubility and only moderate siderophile affinity cause the atmosphere to consistently retain nearly the entire nitrogen budget (2.8 ppmw; Table \ref{tab:volatilebudget}).

\subsubsection{Atmospheres}
\begin{figure}[]
\resizebox{\hsize}{!}{\includegraphics{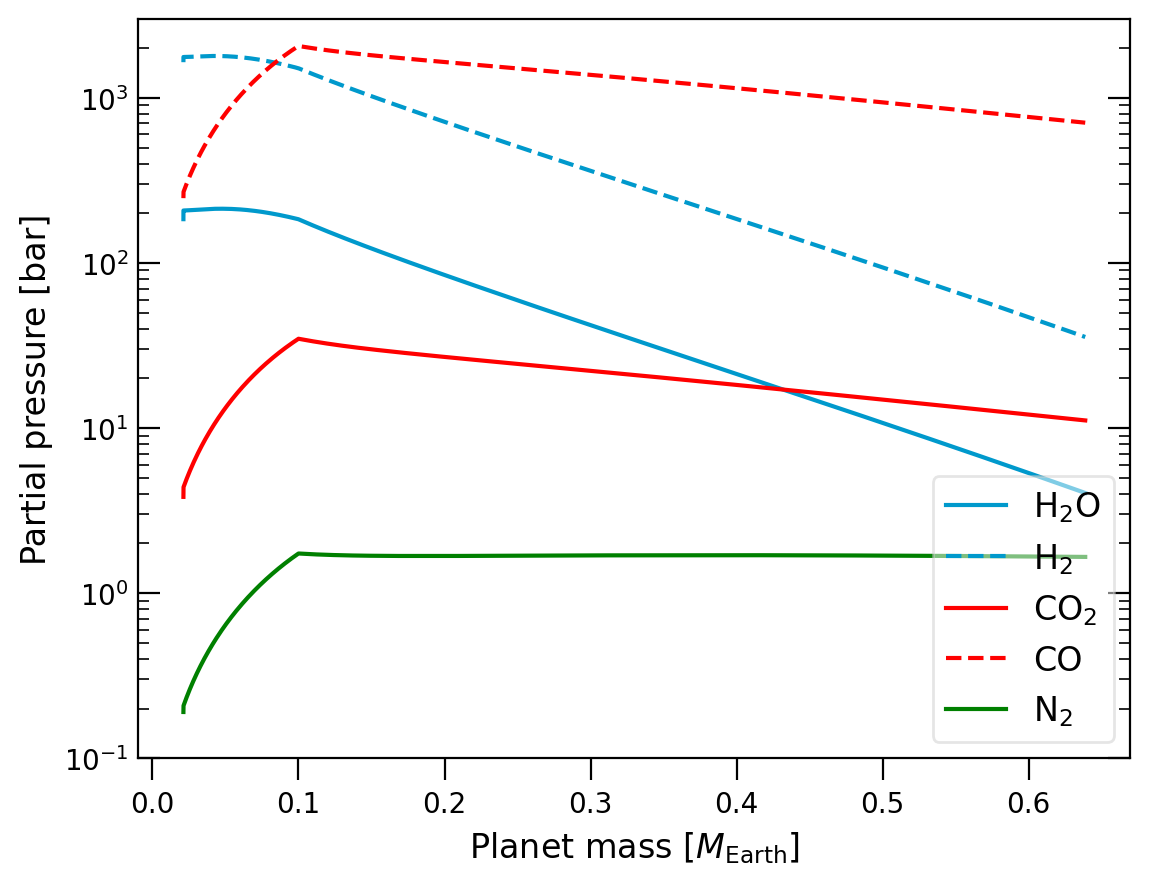}}
\resizebox{\hsize}{!}{\includegraphics{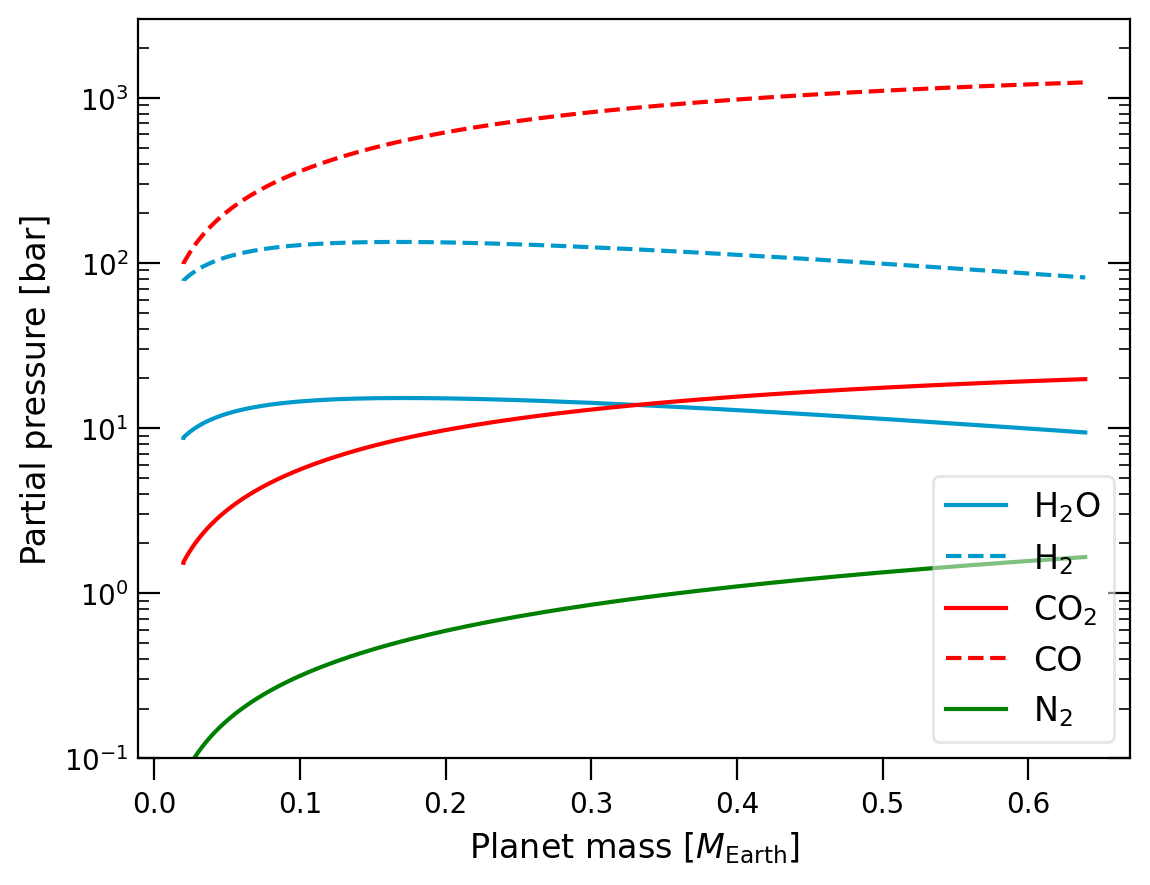}}
\caption{Partial pressure of each volatile-bearing gaseous species as a function of planet mass using constant $\Delta \text{IW} = -2$ (see Eq. \ref{eq:deltaIWspeciation}). Volatile partition coefficients from Luo - Blanchard - Huang. \textit{Top}: The protoplanet starts with the full hydrogen inventory and accretes carbon and nitrogen until $M_\text{p} = 0.1 \, M_\text{E}$, which represents the pebble accretion case (see Sect. \ref{sec:planetgrowth}). The total surface pressure reaches $\sim$$760$ bar at the end of accretion. \textit{Bottom}: The protoplanet accretes all volatiles continuously during its entire growth phase, which represents the planetesimal accretion case. The total surface pressure reaches $\sim$$1350$ bar at the end of accretion.}
\label{piatm}
\end{figure}

Figure \ref{piatm} shows the evolution of the atmosphere as the planet accretes in the Lo: L-B-H model and with early or continuous delivery of volatiles. In the top panel, representing the pebble accretion scenario, hydrogen accretion has already stopped at the initial mass of the protoplanet, but carbon and nitrogen continue to accrete until $M_\text{p} > 0.1 \, M_\text{E}$. After the end of volatile accretion, the partial pressure of each volatile decreases monotonically during the growth of the protoplanet due to continued sedimentation. The partial pressure of nitrogen decreases little between the end of nitrogen delivery and the end of pebble accretion due to inefficient sedimentation into the core compared to both hydrogen and carbon (Fig. \ref{widereservoirpebble}, top row). At the final mass of our pebble-grown proto-Earth analogue, the remaining atmosphere contains roughly 4.0 bar of $\wat$, 35 bar of H$_2$, 11 bar of $\car$, 710 bar of CO, and 1.7 bar of $\nit$.

The bottom panel of Fig. \ref{piatm} shows the atmosphere retrieved from the planetesimal accretion Lo: L-B-H model. In contrast to the corresponding pebble accretion scenario, the partial pressures of both carbon and nitrogen increase during the entire growth of the planetesimal-accreting protoplanet as a result of continuous volatile accretion to the atmosphere and inefficient dissolution. However, the partial pressures of both hydrogen-bearing species, $\wat$ and H$_2$, decrease as the protoplanet grows. This indicates that hydrogen dissolution is sufficiently efficient to outpace the continuous delivery of new hydrogen. The planetesimal accretion model produces higher atmospheric pressures for all volatiles than the corresponding pebble accretion model. The final atmosphere of the planetesimal-grown proto-Earth contains roughly 9.4 bar of $\wat$, 82 bar of H$_2$, 20 bar of $\car$, 1240 bar of CO, and 1.7 bar of $\nit$.

\begin{figure}[!t]
\resizebox{\hsize}{!}{\includegraphics{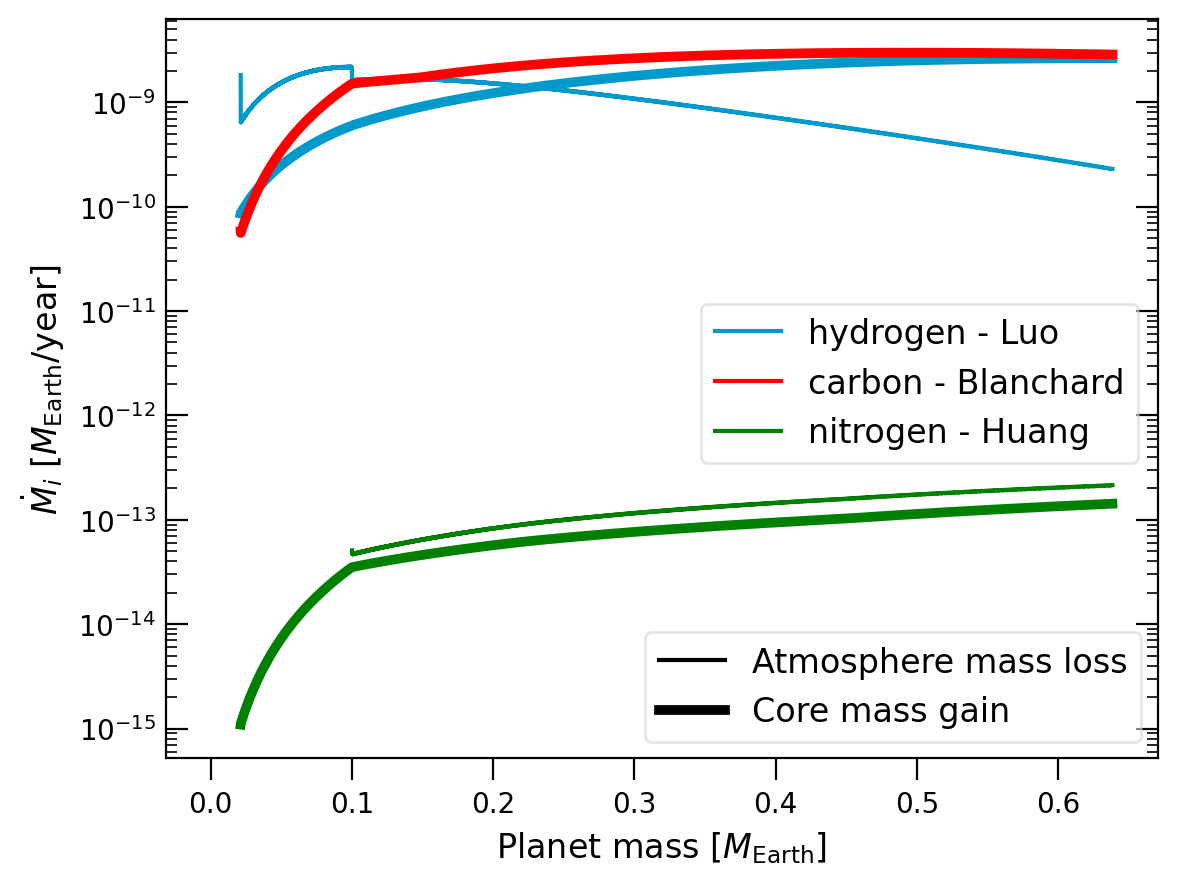}}
\caption{The rate of volatile exchange between planet reservoirs. Thick lines indicate the mass sedimentation rate into the core and thin lines indicate the mass loss rate of the atmosphere, representing net dissolution into the magma ocean. Masses are reported as the mass of the oxidized species of the volatile ($\wat$, $\car$, $\nit$). The atmospheric volatile mass loss rates are only plotted after the volatile has stopped accreting. Prior to $0.1$$M_\text{E}$ the atmosphere continuously accretes both carbon and nitrogen.}
\label{masschangeboth}
\end{figure}

Fig. \ref{masschangeboth} shows the rate at which volatiles move between reservoirs of the planet, from the surface to the magma ocean and from the CMB to the core, in the Lo: L-B-H model. Carbon and nitrogen exhibit strongly coupled atmosphere mass loss rates and core sedimentation rates. For carbon specifically, the atmosphere mass loss rate and the core mass gain are indistinguishable. Mass conservation relates the exchange rates between the planet reservoirs as
\begin{equation}
    \frac{dM_{i}^\text{(core)}}{dt} +  \frac{dM_{i}^\text{(atm)}}{dt} +  \frac{dM_{i}^\text{(mantle)}}{dt} = 0,
\end{equation}
which explains this correlation when $\frac{dM_{i}^\text{(mantle)}}{dt} \approx 0$, as is the case for carbon and nitrogen because they are siderophile and dissolve poorly in the magma ocean. However, hydrogen which dissolves well in magma has a large fraction of its total mass in the magma ocean (Fig. \ref{widereservoirpebble}, top row). Hydrogen dissolves in the magma faster than metal droplets can deliver the dissolved hydrogen to the core. 

\subsection{Effective partition coefficient} \label{sec:effective}
As shown in Sect. \ref{sec:fate}, the mass of volatiles in the core depends on a combination of their partition coefficients and magma-atmosphere solubilities. Both $\wat$ and H$_2$ are more soluble than $\nit$ (Table \ref{tab:solubility}). Consequently, in our pebble accretion Lo: L-B-H model, only 4\% of the nitrogen mass enters the core, whereas 70\% of the hydrogen mass enters the core, despite nitrogen being more siderophile than hydrogen for all $(p, T)$ (see Figs. \ref{partitioncoef} and \ref{widereservoirpebble}, top row). Atmophile elements, such as carbon and nitrogen, store a large fraction of their mass budget in the atmosphere, safe from sedimentation into the core. To quantify this effect and the impact of early volatile delivery, we introduce an effective partition coefficient
\begin{equation}
    D_{i, \text{eff}} = \frac{M_\text{mantle}}{M_\text{core}} \frac{M_{i}^\text{(core)}}{M_{i}^\text{(atm)}+M_{i}^\text{(mantle)}}, \label{Dieff}
\end{equation}
which converges to $D_i$ (Eq. \ref{eq:partitioncoef}) if there was no atmosphere, volatile delivery was continuous, and there was full core-mantle equilibration (see Sect. \ref{sec:magmaoceantransport}). By construction, $D_{i, \text{eff}}$ accounts for both the solubility of volatiles, the integrated history of changes to the partition coefficient as the planet grows, and, in the pebble accretion case, the cumulative effect of early volatile delivery. 

\begin{figure}[]
\resizebox{\hsize}{!}{\includegraphics{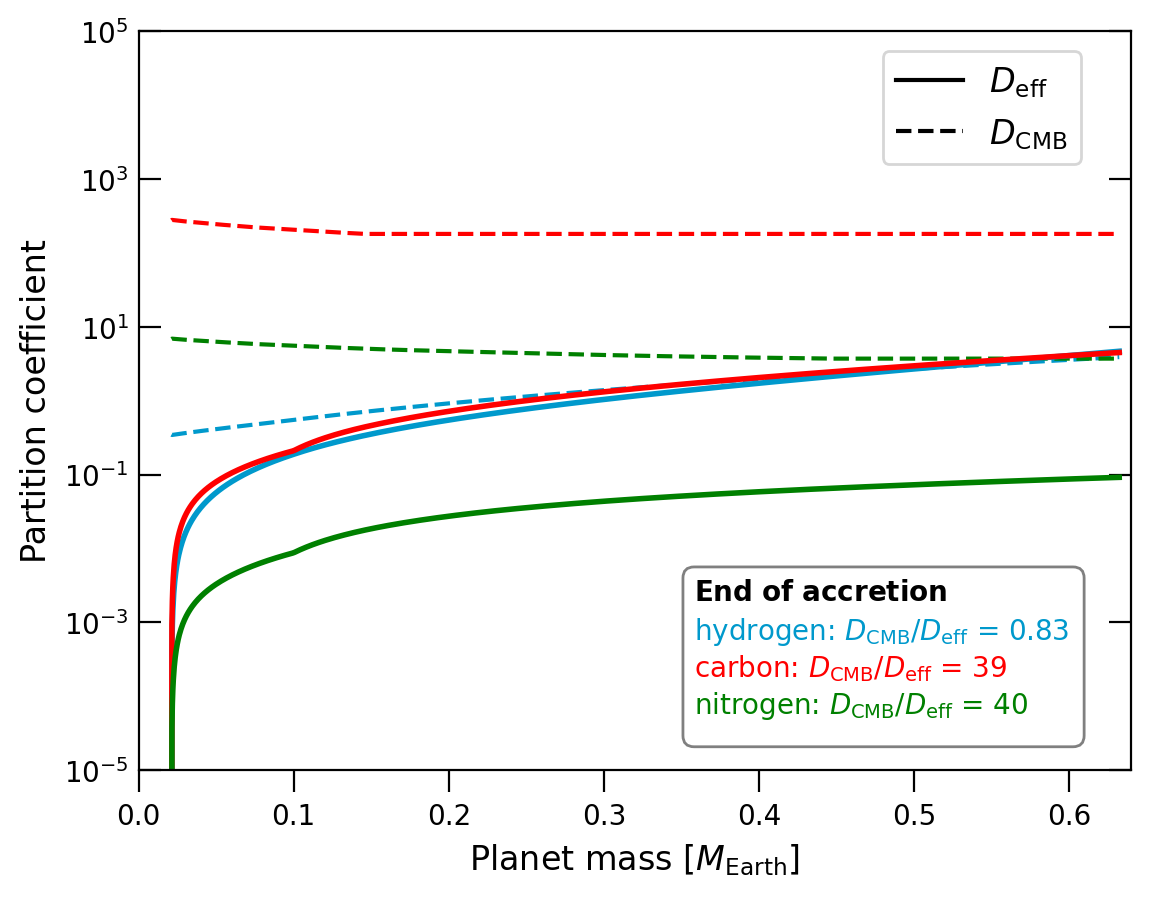}}
\caption{Effective partition coefficient (Eq. \ref{Dieff}) versus partition coefficient at the CMB for a rocky planet growing by pebble accretion and using the volatile partition coefficients from \citet{Luo2024, Blanchard2022} and \citet{Huang2024}. Hydrogen, carbon and nitrogen stop accreting after $M_\text{p} > [0.02, 0.1, 0.1] \, M_\text{E}$, respectively (see Sect. \ref{sec:planetgrowth}). The effective partition coefficient does not necessarily converge to its corresponding CMB value. Box legend shows the relation between the partition coefficient and the effective partition coefficient at the end of accretion.}
\label{PeffcomparisonBlanchard}
\end{figure}

Fig. \ref{PeffcomparisonBlanchard} shows the effective partition coefficient in the Lo: L-B-H pebble accretion model. The effective partition coefficient does not necessarily converge to its value at the surface or at the CMB due to the effect of solubility and the early accretion of volatiles (see Sects. \ref{sec:planetgrowth} and \ref{sec:magmaoceantransport}). The apparent correspondence between the effective partition coefficient of $\wat$ and its value at the CMB is due to a surprising balance between the enhanced sedimentation effect of early volatile delivery and the inhibiting effect of storing hydrogen in the atmosphere. Sedimentation of carbon and nitrogen remains inefficient due to their relative insolubility, causing the effective partition coefficients to differ from the CMB partition coefficients by 1.6 dex. 

\begin{figure}[]
\resizebox{\hsize}{!}{\includegraphics{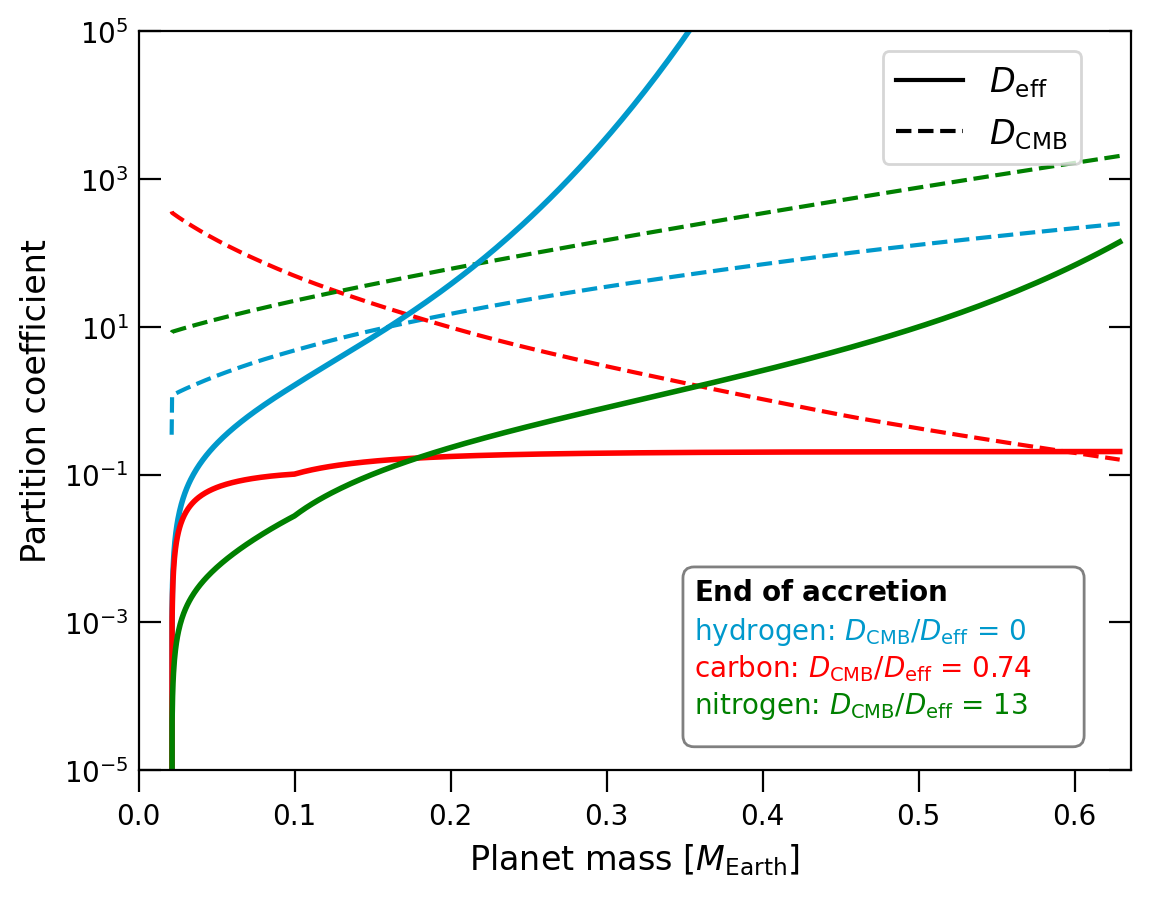}}
\caption{Same as Fig. \ref{PeffcomparisonBlanchard} but for the partition coefficients from \citet{Tagawa2021, Fischer2020} and \citet{Grewal2019}. The effective partition coefficient of hydrogen greatly exceeds the CMB partition coefficient due to the combined effect of early delivery, high solubility, and siderophile affinity (see Eq. \ref{eq:twocomponent}).}
\label{PeffcomparisonFischer}
\end{figure}

Figure \ref{PeffcomparisonFischer} shows the effective partition coefficient as the planet grows using the partition coefficients from \citet{Tagawa2021, Fischer2020} and \citet{Grewal2019}. The effective partition coefficient of hydrogen grows far beyond the naive partition coefficient due to runaway sedimentation because hydrogen is delivered early and dissolves well into the magma ocean (see Sect. \ref{sec:magmaoceantransport}). In contrast, the effective partition coefficient of carbon stops growing because $D_\text{C, CMB}$ decreases as the planet grows. The effective partition coefficient of nitrogen is initially far below the CMB value, due to insolubility limiting its sedimentation. As the planet grows and nitrogen becomes more siderophile, the effective nitrogen partition coefficient increases noticeably.

\section{Discussion} \label{sec:discussion}
The evolution of volatile inventories in planetary reservoirs depends sensitively on the planet's accretion history and on parameters that govern the transport of volatiles. We discuss now how these factors affect our predictions of the volatile inventories of rocky planets at the end of their accretion.
In Sect. \ref{sec:afteraccretion} we describe the thermal and chemical evolution of Earth after the end of its magma ocean phase. In Sect. \ref{sec:theia} we discuss how volatile inventories would change as a result of the Moon-forming giant impact and the effect of volatile delivery timing. In Sect. \ref{sec:earth} we compare our model results with Earth's bulk silicate inventory of volatiles. In Sect. \ref{sec:venus} we detail how our volatile transport model would extrapolate to Venus and reconcile model predictions with present-day observations of Venus' surface inventory of volatiles. In Sect. \ref{sec:exoplanets} we discuss the difficulties in extrapolating our models to higher mass exoplanets in the super-Earth and sub-Neptune regime. Finally, in Sect. \ref{sec:inputparameters} we examine the uncertainties in, and the model sensitivity to, the total volatile inventory of Earth, volatile partition coefficients, solubilities, and delivery timing. 

\subsection{Evolution after accretion has terminated} \label{sec:afteraccretion}
Earth's mantle likely oxidized significantly at the end of accretion \citep[e.g.][]{Trail2011, Kuwahara2023, Zhang2024constant, Zhang2024ferric}. When a planet stops accreting, the flux of fresh metal to the mantle vanishes, greatly reducing the metallic Fe abundance in the magma ocean. 
With little or no metallic Fe remaining, the redox equilibrium (Eq. \ref{eq:FeOtoFeO1.5}) produces more FeO$_{1.5}$ in the deep magma ocean, thereby oxidizing the mantle \citep[e.g.][]{Hirschmann2012, Zhang2017, Armstrong2019, Deng2020, Sossi2020}. Because this production of ferric iron (Fe$^{3+}$) at the end of accretion increases with the depth of the magma ocean, differences between the size of the bodies may explain the relatively oxidized mantles of Earth and Venus and the relatively reduced mantles of Mars, Mercury, Vesta and the Moon \citep[e.g.][]{Armstrong2019, Deng2020, Zhang2024ferric}. 
This oxidizing transition should occur quickly after the end of accretion, on timescales comparable to the $\sim$month-long metal-sinking timescale. Because the magma ocean may not yet have formed a crust during this transition, dissolution will most likely still be ongoing. The oxidized atmosphere's shift toward the more soluble species $\wat$ and $\car$ (Table \ref{tab:solubility}) could then cause further dissolution without subsequent sedimentation to the core.

As the magma ocean crystallizes and cools, oxidized speciation becomes further favored in part because the equilibrium coefficients of the speciation reactions ($K_{\text{H}_2}$, $K_{\text{CO}}$; Eqs. \ref{rec:H2spec}, \ref{rec:COspec}) decrease with temperature \citep[e.g.][]{Ortenzi2020, Sossi2020}. The oxygen fugacity also changes as the surface cools because the activity coefficients of the melt buffer components depend on temperature \citep[e.g.][]{Sossi2020}, which also affects the speciation of volatiles (Eq. \ref{eq:deltaIWspeciation}). Furthermore, at lower temperatures, the water-gas shift reaction
\begin{equation}
    \text{CO~(g)} + \wat\,\text{(g)} \rightleftharpoons \car~\text{(g)} +\text{H}_2\text{(g)}
\end{equation}
increases $\car\,\text{(g)}$ production at the expense of $\wat\,\text{(g)}$ \citep[e.g.][]{Johansen2024}. At even lower temperatures, both H$_2$ and CO may oxidize via precipitation of graphite \citep[e.g.][]{Keppler2019, Sossi2020}
\begin{equation}
    \text{H}_2\,\text{(g)} + 3\,\text{CO~(g)} \rightleftharpoons \car\,\text{(g)} + \wat\,\text{(g)} + 2\,\text{C}\,(\text{s}),
\end{equation}
which also favors transition to a $\wat$ and $\car$-dominated atmosphere. 

As the magma ocean crystallizes bottom-up, a portion of the $\wat$ in the magma ocean may remain trapped between newly formed silicate crystals \citep{Bower2022, Miyazaki2022}. The remaining outgassed $\wat$ would condense only if the ratio $\wat/\car$ is sufficiently high, because $\car$ is a greenhouse gas that can maintain surface temperatures above the sublimation temperature of $\wat$. For our proto-Earth analogue at 1 au, the outgassed $\wat$ likely exceeds the surface water limit required to rapidly condense an ocean imposed by its $\car$ atmosphere \citep[e.g.][]{Salvador2017, Miyazaki2022}. During the condensing of Earth's oceans, hydrogen was vulnerable to atmospheric escape due to the young Sun emitting a high XUV flux \citep[e.g.][]{Watson1981, Kuramoto2013, Kasting1983, Way2020}.

\subsection{Theia and the Moon} \label{sec:theia}
The Moon-forming giant impact, and possibly multiple other giant impacts \citep[e.g.][]{Quintana2016}, further complicates models of the volatile history of Earth. The collision between proto-Earth and its impactor Theia provided sufficient energy to remelt the mantle and allow further exchange of volatiles between the surface and the core \citep[e.g.][]{Sleep2001, Fegley2020, Davis2024} after pebble accretion and planetesimal accretion in the protoplanetary disk phase had already ceased. If Theia formed by pebble accretion, it would have had a higher volatile concentration than proto-Earth because Theia's accretion stopped closer to the mass threshold for volatile sublimation (see Sect. \ref{sec:planetgrowth}). If both proto-Earth and Theia instead formed from planetesimal accretion, and if those planetesimals had identical volatile abundances, one would expect Theia to have contributed an amount of volatiles proportional to its own mass relative to that of Earth. Volatile delivery by Theia would add "late" delivery of volatiles, in contrast to our "early" and "continuous" scenarios. We stress that "early" and "late" are defined by the fraction of core mass already accreted when the volatiles are delivered (Eqs. \ref{eq:twocomponent}, \ref{eq:threecomponent}) and do not relate to an absolute time, such as the timing of the Moon-forming giant impact. The total volatile budgets of our proto-Earth are based on estimates of Earth's reservoirs (see Sect. \ref{sec:planetgrowth}), but because our model neglects the Moon-forming giant impact, we neglect late volatile delivery and may therefore overestimate the sedimentation to the core (see Sect. \ref{sec:magmaoceantransport}). When Theia impacted proto-Earth, the volatiles in Theia's core may, however, not have fully equilibrated because the degree of equilibration depends on the unknown impact latitude \citep{Maas2021} and relative rotation rate \citep{Canup2008}. Theia's volatile contribution to Earth's mantle and atmosphere is thus uncertain. 

The dependence on the timing of volatile delivery causes order unity corrections when comparing early and continuous volatile delivery (Table \ref{tab:BSE}) for models using the L-B-H partition coefficients. In Sect.~\ref{sec:magmaoceantransport} we showed that highly soluble ($\wat$) and siderophile ($\car$) volatiles depend most sensitively on their delivery timing.
Partition coefficients that vary significantly as the planet grows can further increase the dependence of volatile inventories on the timing of volatile delivery. The L-B-H partition coefficients vary only slightly as the planet grows and therefore have a negligible effect on the general dependence on delivery timing. The T-F-G partition coefficients, however, vary significantly. At the end of the accretion of proto-Earth ($M_\text{p} = 0.64 \, M_\text{E}$), the T-F-G partition coefficients at the CMB are $D_{\wat}\approx10^3$, $D_{\text{C}}\approx10^{-2}$, and $D_{\text{N}}\approx10^3$ (Fig. \ref{PeffcomparisonBlanchard}). Because carbon becomes increasingly lithophile at the CMB and unable to sediment in the \citet{Fischer2020} case, any carbon accreted afterward from late planetesimal accretion or giant impacts would therefore enter the atmosphere of the planet if the carbon in the core of the impactor equilibrates \citep[e.g.][]{Sleep2001, Fegley2020, Davis2024}. Water becomes extremely siderophile at higher planet masses in the \citet{Tagawa2021} case, which implies that late planetesimals or giant impactors may mostly contribute to the core content of H. However, lower mass impactors have experienced less volatile sedimentation to their core. For example, impactors with masses below $0.1 \, M_\text{E}$ may store the majority of the water outside the core, even when adopting the highly siderophile water partition coefficient of \citet{Tagawa2021} (see Figs. \ref{widereservoirpebble} and \ref{widereservoirplanetesimal}; bottom rows). If the impactor metal sinks before the protoplanet reaches dissolution equilibrium, then the bulk silicate reservoirs would retain most of the late accreted volatiles. 

\subsection{Anchoring our model to Earth} \label{sec:earth}
Our simulations explore the growth of proto-Earth by varying the formation scenario, volatile partition coefficients, solubilities, total volatile budget, and oxygen fugacity. Even for models sharing the same total volatile budget, the resulting volatile distributions among planetary reservoirs differ significantly between models. In principle, agreement between a model's bulk silicate volatile concentration and that of Earth could provide a test of the consistency of the model; however, as we discuss below and in Sect. \ref{sec:inputparameters}, uncertainties in both the reference BSE concentrations and in the assumed model parameters prevent conclusive inferences. 

We anchor the total volatile budget to estimates of the BSE concentration and Earth's core content of volatiles (see Sect. \ref{sec:planetgrowth}).
Bulk silicate concentrations measure the volatile mass not sequestered in the core, which enables comparison between proto-Earth and the present day. This is because, once the magma ocean crystallizes, subsequent geological evolution generally does not change the BSE concentration, although some elements may slowly diffuse across the CMB into the lower mantle even after crystallization \citep[e.g.][]{Hayden2007, He2023}. In contrast, carbon subduction -- a significant mode of volatile cycling -- transports carbon from the exosphere to the mantle on tectonic timescales \citep{Plank2019}, which reduces the carbon surface inventory, but does not affect the carbon BSE concentration. Therefore, to first order, the present-day BSE likely approximates the volatile bulk silicate concentrations at the end of accretion. 

Several of our models produce bulk silicate concentrations (Table \ref{tab:BSE}) that fall within the range of published present-day BSE estimates \citep{Marty2012, Hirschmann2018, Marty2020}. However, these estimates differ by amounts larger than, or comparable to, the differences between the results of our models. For example, in our fiducial Lo: L-B-H models, pebble accretion produces a $\wat$ bulk silicate concentration that is close to the $\wat$ BSE estimate of \citet{Hirschmann2018}, while the corresponding planetesimal accretion model is more consistent with \citet{Marty2012}. We note that the Lo: L-B-H pebble accretion model is our only model to simultaneously produce bulk silicate concentrations consistent with the range of BSE estimates for all three volatiles. All our models apart from the Lo: L-B-H pebble accretion model yield carbon concentrations in excess of the BSE estimates. However, the Lo: L-B-H planetesimal accretion model is relatively close to the upper uncertainty bound of the C BSE estimate of \citet{Marty2012}. The substantial spread among the BSE estimates therefore limits the extent to which these model comparisons can discriminate between pebble and planetesimal accretion as formation pathways for Earth.

Hybrid accretion scenarios, where Earth accretes significant mass from both pebbles and planetesimals (e.g.~\citealt{Wang2026}), would produce volatile inventories that lie between the results of our two end member scenarios of pure pebble and pure planetesimal accretion.

\subsection{Extension to Venus} \label{sec:venus}
We chose to model a proto-Earth analogue so that the results could be compared to measurements of the volatile inventories of Earth. However, our model can be extended to other rocky planets. Venus, for example, has a slightly larger mass (0.82 $M_\text{E}$) than our proto-Earth (0.64 $M_\text{E}$). Although we did not model volatile transport up to the mass of Venus because of extrapolation concerns, the process of sedimentation continuously dissolving the atmosphere remains as long as the planet accretes material which contains metal but is depleted in volatiles. Given similar volatile abundances of the accreted material, our model predicts an initial atmosphere similar to that of our proto-Earth. Extrapolating the atmosphere to Venus' mass (Fig. \ref{piatm}) yields a $\sim$$500$ bar CO atmosphere prior to atmospheric oxidation (Sect.~\ref{sec:afteraccretion}). However, we also anchored our model's total volatile budget to the estimated inventory of Earth which might differ from the initial inventory of Venus. If Theia indeed contributed a significant portion of Earth's volatiles, and if Venus did not have a comparable contribution from giant impacts or late planetesimal accretion, then this would further reduce our estimate for the mass of Venus' initial atmosphere. 
 
Additionally, the closer orbit of Venus implies two key differences. First, if more of the source material of Venus formed closer to the Sun, potentially inside of the $\wat$ snow-line, then Venus might have had a smaller initial inventory of $\wat$. However, Venus and Earth may have both formed outside the water ice line and subsequently migrated inward \citep[e.g.][]{Anatomy2}. Second, Venus' proximity to the Sun would increase the rate of luminosity-driven volatile loss, mainly H$_2$ escape, compared to that of Earth, causing Venus to lose its initial water inventory \citep[e.g.][]{Kasting1983, Salvador2017, Way2020, Johansen2024}. The observed high D/H ratio of Venus' atmosphere is consistent with significant water loss, although its timing is uncertain \citep{Donahue1982, Grinspoon1993, Gillmann2022, Zahnle2023}. Enhanced D/H ratios may also be produced from metal partitioning favoring the lighter hydrogen isotope \citep{Zhang2026}. Three-dimensional general circulation models support the possibility that Venus retained a shallow water ocean for Gyr timescales before triggering a runaway greenhouse transition to reach its current climate \citep[e.g.][]{Way2020}. The current stagnant lid tectonic mode of Venus also differs from Earth's mobile plate tectonics which allow for the subduction of carbon \citep[e.g.][]{Plank2019, Honing2019, Honing2021}. To enable mobile plate convection, an additional mechanism is required to reduce the yield stress of the lithosphere. Earth's unique mobile plate tectonics, and Venus' lack thereof, may therefore owe to the weakening influence of the higher water abundance in Earth's lithosphere \citep[e.g.][]{Moresi1998, Korenaga2010}. However, both Earth's and Venus' tectonic modes may have been completely different in the past \citep[e.g.][]{Cawood2022, Lourencco2020, Way2020, Marshall2024, Weller2020}. Nevertheless, the distinct tectonic modes of Earth and Venus offer a possible explanation for the difference between their atmospheric inventories of carbon, even if their initial inventories were similar: Venus may have been unable to subduct its carbon rich initial atmosphere.

\subsection{Exoplanets} \label{sec:exoplanets}
Beyond the solar system, most known rocky planets are super-Earths in close orbits. Above super-Earth masses lie the sub-Neptunes: exoplanets with an extensive hydrogen envelope and a possibly volatile-rich rocky interior. Extending our model toward exoplanets in higher mass regimes introduces several problems. In general, the adiabatic temperature gradient with respect to depth is less steep than the solidus and liquidus temperature gradients \citep{Katz2008}. Unlike our low-mass planets, whose entire magma oceans exceed the liquidus even for a fixed surface temperature, high-mass planets with sufficiently deep magma oceans may solidify at the CMB when fixing the surface temperature, thereby preventing further volatile sedimentation into the core. Whether the CMB solidifies in higher mass planets depends on their surface temperatures, which are expected to increase with planetary mass and the mass of their accreted hydrogen-helium envelope \citep[e.g.][]{Olson2022, Steinmeyer2024, Tang2025, Rogers2025}. Sufficiently high surface temperatures could even produce a supercritical magma ocean surface \citep{Caracas2023}. Modeling super-Earth or sub-Neptune magma oceans therefore requires evolving the surface temperature as the planet grows \citep[e.g.][]{Steinmeyer2026}. 

Interactions between the atmosphere and the magma ocean are a key component of our volatile transport model. Although volatile solubility experiments reach adequate pressures, the temperatures tested \citep[e.g.][]{Lichtenberg2021} are much lower than the surface temperatures expected for more massive rocky planets. Furthermore, super-Earths \citep{Rogers2025} and sub-Neptunes \citep{Werlen2025} experience extensive water loss by atmospheric escape and chemical reactions at the magma ocean surface, respectively. If the planet has an envelope whose mass fraction exceeds $10^{-4}$ of the total mass of the planet, water reacts with Si-bearing magmas, metals, and vapors to significantly reduce the water content of the atmosphere and envelope \citep[][]{Werlen2025, Lee2025}. For sub-Neptunes in particular, the presence of an H$_2$ envelope also invalidates certain assumptions of our model. We assume that the IW buffer of the magma ocean mediates the speciation between H$_2$ and $\wat$ with negligible effect on the oxygen content of the magma ocean (Eq. \ref{rec:H2spec}). This assumption is equivalent to requiring that surface hydrogen is negligible compared to magma ocean oxygen, and would be less justified given an extensive H$_2$ envelope in contact with the magma ocean \citep{Werlen2025}. Therefore, extending our model to planets more massive than Earth would require a broader chemical network that explicitly includes Si and O. 

\subsection{Input parameter uncertainties} \label{sec:inputparameters}
Siderophile elements, such as carbon, produce BSE concentrations that are remarkably sensitive to their solubility (see Eqs. \ref{eq:twocomponent}, \ref{eq:threecomponent}). 
Previous studies \citep[e.g.][]{Ortenzi2020, Anatomy3} have commonly neglected CO solubility, because $\car$ is an order of magnitude more soluble than CO (Table \ref{tab:solubility}). This assumption is valid if Earth's mantle was oxidizing during accretion \citep[][]{Badro2015, Anatomy3}. Under reducing conditions, however, CO becomes more abundant than $\car$ by more than an order of magnitude, such that CO dissolution dominates the overall solubility of carbon. In contrast, under the reducing conditions assumed for our model ($\text{IW}$$-2$) the dissolution of $\wat$ dominates that of H$_2$, even if the solubility of H$_2$ adopted in our model \citep{Hirschmann2012solubility} would have been underestimated by a factor of three \citep[see Sect. \ref{sec:dissolution}]{Miozzi2025}.
Of all atmospheric species that we consider, the solubility of CO is the most influential for the model results (Fig. \ref{bubbleBSE}) and also the most experimentally uncertain \citep{Armstrong2015, Hirschmann2016, Yoshioka2019, Lichtenberg2021}. 
Future experiments on CO dissolution at higher temperatures would significantly improve the accuracy of volatile abundance estimates in rocky planet reservoirs. 

Most of the uncertainty in our results comes from the tension between experimental partition coefficient fits to pressure and temperature. Uncertainties propagate from the uncertainties of both the partition coefficients and the volatile solubilities. However, while the partition coefficients have orders of magnitude of tension, different solubility coefficients (mostly) only differ by order unity corrections. A previous study of volatile distribution between planetary reservoirs during planetesimal accretion found H and C sequestration comparable to our continuous L-B-H model, but also predicted much greater N sequestration in the core (69--99.9\%), due to the adoption of a higher solubility and partition coefficient than considered here \citep{Gu2024}. When adopting the more siderophile partition coefficient fit of \citet{Grewal2019}, our continuous accretion model predicts a similar $\sim$$92$\% core sequestration of N (Fig.\,\ref{widereservoirplanetesimal}; bottom row), highlighting the model sensitivity to the choice of partition coefficient. 

Another significant source of uncertainty is the total volatile budget of Earth (see Sect. \ref{sec:planetgrowth}), which we base on inferred abundances in Earth's core \citep{Chen2014, Huang2026} and the BSE \citep{Hirschmann2018}. For hydrogen and carbon, most of the budget derives from estimates of Earth's core, which remain highly uncertain. 
Our models that increase the hydrogen budget but have the same formation scenario and partition coefficients produce order unity differences to the volatile distributions among reservoirs. For example, the Lo: L-B-H pebble model sequesters 70$\%$ of its hydrogen in the core, whereas the Hi: L-B-H pebble model sequesters $44\%$ in the core. Increasing the hydrogen budget reduces sedimentation to the core for all volatiles because an H$_2$-dominated atmosphere has a low mean molecular weight that inhibits dissolution for all species (see Eqs. \ref{eq:gammarelation} and \ref{eq:gamma}). 

Unlike hydrogen, we fixed the total carbon and nitrogen budget for all models. \citet{Chen2014} inferred the presence of $0.3$~wt\% C in Earth's core to reproduce seismic observations. Previous core-carbon estimates span a wide range of $0.2$--$1.0$~wt\% \citep[e.g.][]{McDonough1995, McDonough2003, Wood2013, Dasgupta2013}, and our adopted abundance may therefore slightly overestimate or significantly underestimate the carbon content of Earth. 
If our models had adopted the higher estimates in this range, the resulting bulk silicate concentration of our proto-Earth models would increase.

Earth's global nitrogen inventory is also uncertain. We base the total nitrogen budget of our models on adopting the BSE estimate of 2.8~ppmw \citep{Hirschmann2018} and assuming negligible nitrogen in Earth's core. Although our fiducial partition coefficient fit assumes that nitrogen is moderately siderophile \citep{Huang2024}, we find that little of the total nitrogen sediments to the core, owing to poor magma ocean solubility. Therefore, our proto-Earth models recover a bulk silicate concentration that is consistent with the BSE concentration that we assumed for the total nitrogen budget. 

\section{Conclusions} \label{sec:conclusion}
We have calculated volatile (H, C, N) transport between the atmosphere, magma ocean, and core during the growth of a proto-Earth analogue by either pebble accretion (early delivery of volatiles while proto-Earth was small) or planetesimal accretion (continuous delivery of volatiles), using models that varied the total volatile budget, volatile solubilities, metal-silicate partition coefficients, and oxygen fugacity. This approach yields predictions for the volatile inventories of planet reservoirs at the end of primary accretion during the protoplanetary disk phase. 

\begin{enumerate}
    \item Volatile partitioning between the core and the bulk silicate reservoir of protoplanets depends on its formation pathway, via the timing of volatile delivery (Table \ref{tab:BSE}). Early delivery followed by continued growth of the mantle and core significantly increases volatile sedimentation to the core in pebble accretion scenarios (Sect. \ref{sec:magmaoceantransport}). 
    Volatile sedimentation efficiency is particularly sensitive to delivery timing for volatiles that are highly soluble ($\wat$) or highly siderophile (C). Early delivery leads to  $\sim$$70$\% of H and C entering the core, compared to $\lesssim$$50$\% sequestration for both volatiles under continuous delivery (Figs. \ref{widereservoirpebble} and \ref{widereservoirplanetesimal}; top rows). The mantle and atmospheric inventories in the continuous delivery model are therefore approximately twice those of the early delivery model.
    \\
    \item Insolubility limits volatile sedimentation to the core (Figs. \ref{bubbleBSE} and \ref{masschangeboth}). Carbon and nitrogen are sufficiently atmophile to store a significant fraction of their mass in the atmosphere, reducing core sequestration despite their siderophile nature (Figs. \ref{widereservoirpebble} and \ref{widereservoirplanetesimal}). In contrast, the much higher solubility of water allows substantial hydrogen to reside in the magma ocean, causing efficient sedimentation to the core. \\
    \item Metal-silicate partition coefficients fits based on experiments and simulations vary significantly (Fig.\,\ref{partitioncoef}). Adopting different partition coefficient fits between our proto-Earth models produces estimates of the volatile abundances of the bulk mantle and atmosphere that differ by several orders of magnitude (Table \ref{tab:BSE}), demonstrating that our results are most sensitive to the adopted partition coefficients. \\
    \item Volatile solubility and delivery timing affect the cumulative partitioning of volatiles between the core and the bulk silicate reservoir. We calculate effective partition coefficients (final volatile concentration of the core relative to final volatile concentration of the mantle including the atmosphere reservoir) that differ significantly from evaluating partitioning equilibrium between the core and the mantle reservoirs at the final deep mantle pressure 
    (Figs. \ref{PeffcomparisonBlanchard} and \ref{PeffcomparisonFischer}). \\
    \item When adopting our fiducial set of partition coefficients, the primordial atmosphere of proto-Earth at the end of accretion is carbon-dominated, but also contains substantial hydrogen and $\sim$$2$ bar of $\nit$. The pebble accretion model produces a $\sim$$760$ bar atmosphere while the planetesimal accretion model produces a more massive $\sim$$1350$ bar atmosphere (Fig. \ref{piatm}). 
\end{enumerate}

Our volatile transport models predict substantially different volatile inventories in the pebble and planetesimal accretion scenarios, differences that in principle could be used to infer Earth's formation pathway. However, estimates of Earth's present-day volatile inventories vary widely, such that our fiducial pebble and planetesimal accretion models both produce volatile inventories within the range of published estimates. 
Uncertainties in volatile partition coefficients and solubilities also limit discrimination between formation pathways, as increased core sequestration could result from higher partition coefficients or solubilities rather than early volatile delivery.
Further constraints on Earth's volatile inventories, as well as on volatile partition coefficients and solubilities, may, therefore, provide insight into Earth's dominant formation pathway.

\begin{acknowledgements}
V.~S. thanks Claudia Danti and Christian Granzow for endless support and fruitful discussions. V.~S thanks Ayumu Kuwahara for discussion about the structure of the paper. V.~S thanks Geoffrey Andama for assistance in integrating volatile sedimentation to the core. V.~S further thanks Haiyang Wang for discussions on magma ocean dissolution and solubility. V.~S thanks Ziyan Xu for insight on volatile mixing by convection. V.~S thanks Paolo Sossi for helpful discussion and for providing insight into diamond anvil cell partition coefficient studies. V.~S thanks James Badro for discussion regarding the composition of Earth's mantle. The authors acknowledge Adrien Houge for illustrating the schematic overview of the volatile transport model. A.~J. acknowledges funding from the Carlsberg Foundation (Semper Ardens: Advance grant FIRSTATMO).
\end{acknowledgements}

\bibliographystyle{aa} 


\begin{appendix} 
\section{Volatile sedimentation, early delivery} 
\label{sec:twothreecomponentderivation}
We here integrate volatile sedimentation to the core in cases with early delivery of volatiles. If the planet continues to accrete metal and silicates, but no longer accretes volatiles then the volatile concentration of the mantle will decrease over time due to sedimentation to the core. 

As the core grows, the volatile fraction of the sedimented metal is sequestered in the core according to the metal-silicate partition coefficient, 
\begin{equation}
    \frac{d M_\text{i}^\text{(core)}}{d M_\text{core}} = D_i C_i^\text{(sil)}.
\end{equation}
We assume that the mantle concentration is at partitioning equilibrium,
\begin{equation}
    C_i^\text{(sil)} = \frac{M_i^{(\text{mantle})}}{M_\text{mantle}},
\end{equation}
such that
\begin{equation}
    \frac{d M_i^\text{(core)}}{d M_\text{p}} = D_i \frac{M_\text{core}}{M_\text{mantle}} \frac{M_{i}^{(\text{mantle})}}{M_\text{p}}. \label{eq:twothreecomponentstartingpoint}
\end{equation}
where the core-mantle fraction $\frac{M_\text{core}}{M_\text{mantle}}$ is assumed constant.
Equation \ref{eq:twothreecomponentstartingpoint} yields a separable differential equation if one can express $M_i^\text{(mantle)}$ as a function of $M_i^\text{(core)}$.

The total volatile mass is constant after the end of volatile accretion and is the sum of the volatile inventories,
\begin{equation}
    M_\text{i}^{(\text{tot})} = M_\text{i}^{(\text{core})} + M_\text{i}^{(\text{mantle})} + 
    M_\text{i}^{(\text{atm})}.
\end{equation}
In the case where negligible volatile mass is stored in the atmosphere, 
\begin{equation}
    M_\text{i}^{(\text{mantle})} = M_\text{i}^{(\text{tot})} - M_\text{i}^{(\text{core})},
\end{equation}
one can rewrite Eq. \ref{eq:twothreecomponentstartingpoint} as 
\begin{equation}
    \frac{d M_i^\text{(core)}}{M_\text{i}^{(\text{tot})} - M_\text{i}^{(\text{core})}} = D_i \frac{M_\text{core}}{M_\text{mantle}} \frac{dM_\text{p}}{M_\text{p}}
\end{equation}
We integrate both sides to solve for $M_i^\text{(core)}$ as a function of $M_\text{p}$,
\begin{align}
    & \left[ -\ln \, (M_{i}^\text{(tot)} - M_{i}^\text{(core)}) \right]^{M_{i, 1}^\text{(core)}}_{M_{i, 0}^\text{(core)}} = \frac{D_iM_\text{core}}{M_\text{mantle}} \Big[ \ln M_\text{p} \Big]^{M_\text{p, 1}}_{M_\text{p, 0}}, \\
    & \ln \, \left(\frac{M_{i}^\text{(tot)} - M_{i,0}^\text{(core)}}{M_{i}^\text{(tot)} - M_{i,1}^\text{(core)}}\right) = \frac{D_iM_\text{core}}{M_\text{mantle}} \ln \frac{M_\text{p,1}}{M_\text{p,0}},
\end{align}
where $M_{i.0}^\text{(core)}, M_{i.1}^\text{(core)}$ are the volatile masses in the core when the planet masses are $M_\text{p,0}$, $M_\text{p,1}$ respectively. Both $M_\text{p,0}$ and $M_\text{p,1}$ must exceed the planet mass required for sublimation of the volatile $i$. Therefore, to model all of the volatile sedimentation after delivery ceases, we set $M_\text{p,0}$ to be this critical mass for volatile sublimation in the envelope.
We exponentiate both sides,
\begin{equation}
    \frac{M_{i}^\text{(tot)} - M_{i,0}^\text{(core)}}{M_{i}^\text{(tot)} - M_{i,1}^\text{(core)}} = \left(\frac{M_\text{p,1}}{M_\text{p,0}} \right)^{\frac{D_iM_\text{core}}{M_\text{mantle}}},
\end{equation}
then invert both sides and substitute back $M_i^\text{(mantle)}$,
\begin{equation}
    M_{i,1}^\text{(mantle)} = M_{i,0}^\text{(mantle)}\left(\frac{M_\text{p,0}}{M_\text{p,1}} \right)^{\frac{D_iM_\text{core}}{M_\text{mantle}}}. \label{eq:twocomponenttrue}
\end{equation}
In the event that the entire volatile mass is delivered before substantial sedimentation, $M_{i,0}^\text{(mantle)} = M_i^\text{(tot)}$ is the total volatile budget, which yields Eq. \ref{eq:twocomponent} presented in the main text (Sect. \ref{sec:magmaoceantransport}).

If substantial volatile mass is stored in the atmosphere, $M_i^\text{(mantle)}$ is composed of three reservoir terms, 
\begin{equation}
    M_i^{(\text{mantle})} = M_i^{(\text{tot})} - M_i^{(\text{core})} - M_i^\text{(atm)}, \label{eq:mantleinthreecomponents}
\end{equation}
such that direct substitution into Eq. \ref{eq:twothreecomponentstartingpoint} does not produce a separable differential equation. Instead one must first relate $M_i^\text{(atm)}$ to $M_i^\text{(mantle)}$ via dissolution. 
At equilibrium, the partial pressure of a volatile is proportional to the dissolved mass fraction (Eq. \ref{eq:Licht}). However, partial pressure relates to the atmospheric mole fraction, 
\begin{equation}
    p_i = p_\text{surf} \, x_i^\text{(atm)}, 
\end{equation}
which necessitates conversion from mole fraction $x_i^\text{(atm)}$ to mass fraction by the mean molecular weight $\bar{\mu}$ and molecular weight of the volatile $\mu_i$, 
\begin{equation}
    x_i^\text{(atm)} = \frac{M_i^\text{(atm)}}{M_\text{atm}} \frac{\bar{\mu}}{\mu_i}.
\end{equation}
In the thin atmosphere approximation with a negligible envelope pressure, 
\begin{equation}
    p_\text{surf} = \frac{GM_\text{p}M_\text{atm}}{4\pi r_\text{surf}^4}. \label{eq:thinatmosphere}
\end{equation}
We can then express $M_i^\text{(atm)}$ as a function of $p_i$, 
\begin{equation}
    M_i^\text{(atm)} = p_i \frac{4\pi r_\text{surf}^4}{G M_\text{p}} \frac{\mu_i}{\bar{\mu}} =  p_i \left(\frac{3^4 M_\text{p}}{4 \pi G^3\bar{\rho}^4}\right)^{1/3} \frac{\mu_i}{\bar{\mu}} 
\end{equation}
where $\bar{\rho}$ is the mean density of the planet. Substituting $p_i$ for the $M_i^\text{(mantle)}$ at dissolution equilibrium (Eq. \ref{eq:Licht}) yields
\begin{equation}
    M_i^\text{(atm)} = \left({M_i^\text{(mantle)}}\right)^{\beta_i} M_\text{p}^\frac{1-3\beta_i}{3} \frac{1}{G (1-f_\text{met})^{\beta_i} \alpha_i^{\beta_i}} \left(\frac{3^4}{4\pi \bar{\rho}^4}\right)^{1/3}
\end{equation}
from which we define a constant $\gamma_i$,  
\begin{equation}
    \gamma_i = \frac{1}{G (1-f_\text{met})^{\beta_i} \alpha_i^{\beta_i}} \left(\frac{3^4}{4\pi \bar{\rho}^4}\right)^{1/3}
\end{equation}
such that
\begin{equation}
    M_i^\text{(atm)} = \left({M_i^\text{(mantle)}}\right)^{\beta_i} M_\text{p}^\frac{1-3\beta_i}{3} \gamma_i.
\end{equation}
We can substitute this expression into Eq. \ref{eq:mantleinthreecomponents},
\begin{equation}
    M_\text{i}^{(\text{mantle})} = M_\text{i}^{(\text{tot})} - M_\text{i}^{(\text{core})} - \left({M_i^\text{(mantle)}}\right)^{\beta_i} M_\text{p}^\frac{1-3\beta_i}{3} \gamma_i,
\end{equation}
to solve for $M_\text{i}^{(\text{mantle})}$, which is only possible for certain values of the dissolution stoichiometry coefficient $\beta_i$. If the solubility of the volatile is Henrian ($\beta_i = 1$), then we get 
\begin{equation}
    M_i^\text{(mantle)} = \frac{M_i^{\text{(tot)}}-M_i^\text{(core)}}{1+\gamma_i M_\text{p}^{-2/3}}.
\end{equation}
Substitution into Eq. \ref{eq:twothreecomponentstartingpoint} yields
\begin{equation}
    \frac{d M_i^\text{(core)}}{M_\text{i}^{(\text{tot})} - M_\text{i}^{(\text{core})}} = D_i \frac{M_\text{core}}{M_\text{mantle}} \frac{dM_\text{p}}{M_\text{p}+\gamma_i M_\text{p}^{1/3}},
\end{equation}
which can be integrated by variable substitution $u = M_\text{p}^{2/3}+\gamma_i$, $du = \frac{2}{3}M_\text{p}^{-1/3}$,
\begin{equation}
    \int \frac{d M_i^\text{(core)}}{M_\text{i}^{(\text{tot})} - M_\text{i}^{(\text{core})}} = D_i \frac{M_\text{core}}{M_\text{mantle}} \int \frac{3}{2}\frac{du}{u},
\end{equation}
to find the anti-derivatives. Thus, 
\begin{align}
    & \left[ -\ln \, (M_{i}^\text{(tot)} - M_{i}^\text{(core)}) \right]^{M_{i, 1}^\text{(core)}}_{M_{i, 0}^\text{(core)}} = \frac{3}{2}\frac{D_iM_\text{core}}{M_\text{mantle}} \Big[ \ln \left(M_\text{p}^{2/3}+\gamma_i\right) \Big]^{M_\text{p, 1}}_{M_\text{p, 0}}, \\
    & \ln \, \left(\frac{M_{i}^\text{(tot)} - M_{i,0}^\text{(core)}}{M_{i}^\text{(tot)} - M_{i,1}^\text{(core)}}\right) = \frac{3}{2} \frac{D_i M_\text{core}}{M_\text{mantle}} \ln \left(\frac{M_\text{p,1}^{2/3}+\gamma_i}{M_\text{p,0}^{2/3}+\gamma_i} \right).
\end{align}
As with Eq. \ref{eq:twocomponenttrue}, we exponentiate to solve for the volatile mass not sequestered in the core ($M_i^{\text{(BSE)}} = M_i^{\text{(mantle)}} + M_i^{\text{(atm)}}$), only now this includes the mass in the atmosphere.
\begin{align}
    & \frac{M_{i,0}^\text{(mantle)} + M_{i,0}^\text{(atm)}}{M_{i,1}^\text{(mantle)} + M_{i,1}^\text{(atm)}} = \left(\frac{M_\text{p,1}^{2/3}+\gamma_i}{M_\text{p,0}^{2/3}+\gamma_i} \right)^{\frac{3}{2} \frac{D_i M_\text{core}}{M_\text{mantle}}}, \\
    & M_{i,1}^\text{(mantle)} + M_{i,1}^\text{(atm)} = \left(M_{i,0}^\text{(mantle)} + M_{i,0}^\text{(atm)} \right)\left(\frac{M_\text{p,0}^{2/3}+\gamma_i}{M_\text{p,1}^{2/3}+\gamma_i} \right)^{\frac{3}{2} \frac{D_i M_\text{core}}{M_\text{mantle}}}. 
\end{align}
If the entire volatile mass is delivered before substantial sedimentation to the core, then $M_{i,0}^\text{(mantle)} + M_{i,0}^\text{(atm)} = M_i^\text{(tot)}$ is the total volatile budget, which yields Eq. \ref{eq:threecomponent} presented in the main text (Sect. \ref{sec:magmaoceantransport}).

\section{Volatile sedimentation, continuous delivery} \label{sec:sedimentationcontinuous}
We here derive volatile sedimentation to the core in the case of continuous delivery of volatiles. If volatiles are continuously delivered onto the protoplanet with a constant volatile mass fraction $f_i$, the total mass of volatile $i$ scales linearly with planet mass,
\begin{equation}
    M_i^\text{(tot)} =  f_i M_\text{p}.
\end{equation}
Consider the case having negligible volatile mass in the atmosphere such that
\begin{equation}
    M_i^\text{(mantle)} = M_i^\text{(tot)} - M_i^\text{(core)} = f_i M_\text{p} - M_i^\text{(core)}.
\end{equation}
Thus, the differential equation of sedimentation (Eq. \ref{eq:twothreecomponentstartingpoint}) becomes
\begin{equation}
    \frac{d M_i^\text{(core)}}{d M_\text{p}} = D_i \frac{M_\text{core}}{M_\text{mantle}} \frac{f_i M_\text{p} - M_i^\text{(core)}}{M_\text{p}}, \label{eq:twocomponentplanetesimaldiff}
\end{equation}
for which $M_i^\text{(core)}$ and $M_\text{p}$ are not immediately separable. We rewrite Eq. \ref{eq:twocomponentplanetesimaldiff} in a standard linear differential equation form as 
\begin{align}
    & \frac{d M_i^\text{(core)}}{d M_\text{p}} + D_i \frac{M_\text{core}}{M_\text{mantle}}  \frac{M_i^\text{(core)}}{M_\text{p}} = D_i \frac{M_\text{core}}{M_\text{mantle}} f_i, \\
    & \frac{d M_i^\text{(core)}}{d M_\text{p}} + P M_i^\text{(core)} = Q
\end{align}
where $P$ and $Q$ are functions of $M_\text{p}$, (Q is a constant function). We multiply both sides of the equations with an integrating factor $U$ (also a function of $M_\text{p}$), defined such that 
\begin{equation}
    \frac{d \left(U M_i^\text{(core)}\right)}{dM_\text{p}} = U\frac{d M_i^\text{(core)}}{d M_\text{p}} + U P M_i^\text{(core)}. \label{eq:integratingfactorproductrule}
\end{equation}
The product rule of derivatives yields
\begin{equation}
    M_i^\text{(core)} \frac{dU}{dM_\text{p}}= U P M_i^\text{(core)}.
\end{equation}
Therefore, $U$ is defined as
\begin{align}
    &U = \exp{\left( \int P \, dM_\text{p} \right)}, \\
    &U = \exp{\left( \int D_i \frac{M_\text{core}}{M_\text{mantle}}  \, \frac{dM_\text{p}}{M_\text{p}} \right)}, \\
    &U = M_\text{p}^{\frac{D_i M_\text{core}}{M_\text{mantle}}}.
\end{align}
By Eq. \ref{eq:integratingfactorproductrule}, the differential equation can thus be rewritten as
\begin{equation}
    \frac{d \left(U M_i^\text{(core)}\right)}{dM_\text{p}} = UQ,
\end{equation}
and integrated to solve for $M_i^\text{(core)}$,
\begin{align}
    & \left[ U M_i^\text{(core)} \right] = \int UQ \, dM_\text{p}, \\
    & \left[ M_\text{p}^{\frac{D_i M_\text{core}}{M_\text{mantle}}} M_i^\text{(core)} \right] = \int D_i \frac{M_\text{core}}{M_\text{mantle}} f_i M_\text{p}^{\frac{D_i M_\text{core}}{M_\text{mantle}}} \, dM_\text{p}, \\
    & \left[ M_\text{p}^{\frac{D_i M_\text{core}}{M_\text{mantle}}} M_i^\text{(core)} \right] = \frac{D_i \frac{M_\text{core}}{M_\text{mantle}}}{\frac{D_i M_\text{core}}{M_\text{mantle}}+1} f_i \left[ M_\text{p}^{\frac{D_i M_\text{core}}{M_\text{mantle}}+1} \right], \\
    & \left[ M_\text{p}^{\frac{D_i M_\text{core}}{M_\text{mantle}}} M_i^\text{(core)} \right] = \frac{D_i M_\text{core}}{D_i M_\text{core}+M_\text{mantle}} f_i \left[ M_\text{p}^{\frac{D_i M_\text{core}}{M_\text{mantle}}+1} \right].
\end{align}
Plugging in the upper and lower bounds yields
\begin{align}
\begin{split}
    &M_\text{p,1}^{\frac{D_i M_\text{core}}{M_\text{mantle}}} M_{i,1}^\text{(core)} - M_\text{p,0}^{\frac{D_i M_\text{core}}{M_\text{mantle}}} M_{i,0}^\text{(core)} = \\
    &= \frac{D_i M_\text{core}}{D_i M_\text{core}+M_\text{mantle}} f_i \left(M_\text{p,1}^{\frac{D_i M_\text{core}}{M_\text{mantle}}+1} - M_\text{p,0}^{\frac{D_i M_\text{core}}{M_\text{mantle}}+1} \right) ,
\end{split}
\end{align}
from which we can solve for the final volatile mass in the core,
\begin{align}
\begin{split}
    &M_{i,1}^\text{(core)} =  \left(\frac{M_\text{p,0}}{M_\text{p,1}}\right)^{\frac{D_i M_\text{core}}{M_\text{mantle}}} M_{i,0}^\text{(core)} + \\
    &+ \frac{D_i M_\text{core}}{D_i M_\text{core}+M_\text{mantle}} f_i \left(M_\text{p,1} - M_\text{p,0} \left(\frac{M_\text{p,0}}{M_\text{p,1}}\right)^\frac{D_i M_\text{core}}{M_\text{mantle}} \right),
\end{split}    
\end{align}
also substituting the definition of constant $f_i = M_i^\text{(tot)}/M_\text{p}$,
\begin{align}
\begin{split}
    &M_{i,1}^\text{(core)} =  \left(\frac{M_\text{p,0}}{M_\text{p,1}}\right)^{\frac{D_i M_\text{core}}{M_\text{mantle}}} M_{i,0}^\text{(core)} + \\
    &+ \frac{D_i M_\text{core}}{D_i M_\text{core}+M_\text{mantle}} \left(M_\text{i,1}^\text{(tot)} - M_\text{i,0}^\text{(tot)} \left(\frac{M_\text{p,0}}{M_\text{p,1}}\right)^\frac{D_i M_\text{core}}{M_\text{mantle}} \right).
\end{split}    
\end{align}
Separately grouping initial and final volatile mass terms yields
\begin{align}
\begin{split}
    &M_{i,1}^\text{(core)} = \frac{D_i M_\text{core}}{D_i M_\text{core}+M_\text{mantle}} M_\text{i,1}^\text{(tot)} + \\
    &+ \left( M_{i,0}^\text{(core)} - M_\text{i,0}^\text{(tot)} \frac{D_i M_\text{core}}{D_i M_\text{core}+M_\text{mantle}} \right) \left(\frac{M_\text{p,0}}{M_\text{p,1}}\right)^{\frac{D_i M_\text{core}}{M_\text{mantle}}}
    \label{eq:coresteadystateconvergence}
\end{split}
\end{align}
for the volatile mass in the core.
The volatile mass in the mantle is expressed similarly,
\begin{align}
\begin{split}
    &M_{i,1}^\text{(mantle)} = \frac{M_\text{mantle}}{D_i M_\text{core}+M_\text{mantle}} M_\text{i,1}^\text{(tot)} + \\
    &+ \left( M_{i,0}^\text{(mantle)} - M_\text{i,0}^\text{(tot)} \frac{M_\text{mantle}}{D_i M_\text{core}+M_\text{mantle}} \right) \left(\frac{M_\text{p,0}}{M_\text{p,1}}\right)^{\frac{D_i M_\text{core}}{M_\text{mantle}}}. \label{eq:mantlesteadystateconvergence}    
\end{split}
\end{align}
Equations \ref{eq:coresteadystateconvergence} and  \ref{eq:mantlesteadystateconvergence} describe convergence toward the steady-state solution derived previously \citep{Anatomy2}. In both equations, the first term on the right-hand side is the steady-state solution, while the second term represents the departure from global partitioning equilibrium in the initial conditions. For any initial condition, the volatile mass in the core approaches the steady-state solution because $\left(\frac{M_\text{p,0}}{M_\text{p,1}}\right)^{\frac{D_i M_\text{core}}{M_\text{mantle}}}$ decreases as the planet grows. 

\end{appendix}

\end{document}